\documentclass[twocolumn,aps,pra,showpacs,superscriptaddress,notitlepage]{revtex4-1}
\usepackage[a4paper,top=3cm,bottom=3cm,left=1.5cm,right=1.5cm]{geometry}
\usepackage[english]{babel}
\usepackage{graphicx}				
\usepackage{booktabs}			
\usepackage{amsmath}		
\usepackage{amsfonts}		
\usepackage{mathtools}				
\usepackage{braket}					
\usepackage{multirow}
\usepackage{setspace}
\usepackage[separate-uncertainty=true]{siunitx}
\usepackage[dvipsnames]{xcolor}
\usepackage{hyperref}
\usepackage{enumitem}
\usepackage{lipsum}

\newcommand{\oos}{\omega_s}
\newcommand{\ooi}{\omega_i}

\newcommand{\ooa}{\left( \oos,\ooi\right)}

\makeatletter
\let\cat@comma@active\@empty
\makeatother

\begin{document}
\title{Design Considerations for High-purity Heralded Single Photon Sources
}

\author{Francesco Graffitti}
\email{fraccalo@gmail.com} 
\affiliation{Scottish Universities Physics Alliance (SUPA), Institute of Photonics and Quantum Sciences, School of Engineering and Physical Sciences, Heriot-Watt University, Edinburgh EH14 4AS, UK}

\author{J\'er\'emy Kelly-Massicotte}
\affiliation{Perimeter Institute for Theoretical Physics, Waterloo, Ontario, N2L 2Y5, Canada}
\affiliation{Department of Physics \& Astronomy, University of Waterloo, Waterloo, Ontario, Canada, N2L 3G1}
\author{Alessandro Fedrizzi}
\affiliation{Scottish Universities Physics Alliance (SUPA), Institute of Photonics and Quantum Sciences, School of Engineering and Physical Sciences, Heriot-Watt University, Edinburgh EH14 4AS, UK}

\author{Agata M. Bra\'nczyk}
\affiliation{Perimeter Institute for Theoretical Physics, Waterloo, Ontario, N2L 2Y5, Canada}

\begin{abstract}
When building a parametric downconversion photon-pair source with spectrally separable photons, e.g. for making high-purity heralded single photons, two practical issues must be accounted for: the design of the experiment, and its characterization. To address experiment design, we study the impact on spectral separability of realistic (sech shaped and chirped) pump fields, realistic nonlinear crystals with fabrication imperfections, and undesirable PDC generation far from the central PMF peak coming from nonlinearity shaping methods. To address experiment characterization, we study the effect of discretization and spectral range of the measured bi-photon joint spectrum, the difference between inferring separability from the joint spectral amplitude vs. the joint spectral intensity, and advantages of interference experiments for purity characterization over methods based on the joint spectral intensity. This study will be of practical interest to researchers building the next generation of nonlinear sources of separable photon pairs.
\end{abstract}

\maketitle

\section{Introduction}
Photon pairs generated by parametric downconversion (PDC) form the backbone of many quantum optics experiments. Photon pairs can be used directly, or as a resource for heralded single photons. In either case, the success of such experiments relies on the quality of the generated PDC photon pairs. 

Of particular importance are the photons' spectral and temporal properties, captured by the \emph{joint spectral amplitude} (JSA). In some  cases, correlations in the JSA are desirable, but more often than not, they are problematic and ought to be minimized. For example, in a heralded single-photon-source, a separable JSA ensures high-purity single photons, which are  necessary for high-visibility interference in optical networks. When both photons of a generated pair are used, either in interference experiments, or as polarization-entangled qubits, a separable JSA is also preferred to avoid contamination during interference. 

The PDC process involves shining a classical pump field onto a nonlinear crystal.  Two practical issues must be accounted for when building a source of spectrally separable photons. The first is the design of the pump and crystal properties. The second is the characterization of the photons' joint spectrum. In this paper, we address both the experiment design and characterization. We focus on practical aspects that have so far been overlooked or, in some cases,  treated incorrectly.

Pump spectrum design for generation of uncorrelated photon pairs is a well-studied problem \cite{grice1997spectral,keller1997theory,erdmann2000restoring,grice2001eliminating,kim2002generation,giovannetti2002extended,giovannetti2002generating,u2006generation,christ2009pure,kaneda2016heralded,weston2016efficient,laudenbach2017numerical,barbieri2017hong}. In virtually all studies, however, the pump laser was taken to have a transform-limited Gaussian spectral amplitude or delta-function distribution; when in reality, a pulsed laser has a sech-shaped spectral amplitude that may not be transform limited. We show how a sech shaped pump and a non-transform limited (chirped) pump impacts the JSA separability. 
 
The design of crystal properties for generation of uncorrelated photon pairs is also a well-studied problem.
This typically involves matching the group velocities of the fields inside the crystal  \cite{grice2001eliminating,u2006generation,mosley2008conditional,mosley2008heralded}, as well as shaping the crystal's nonlinearity profile to approximate a Gaussian function \cite{dixon2013spectral,dosseva2016shaping,tambasco2016domain,graffitti2017pure}. Previous work on tailoring crystal nonlinearities for JSA separability assumed ideal crystal fabrication, and for the most part neglected discussion of undesirable PDC generation that arises from nonlinearity shaping methods. We show how fabrication imperfections in nonlinearity shaping  impact JSA separability and pair generation probability, and discuss the implications of undesired PDC generation. 

Finally, we address the question of how PDC photon sources have been characterised for estimation of properties such as photon purity. The spectral separability of PDC photons can be inferred by measuring the JSA at discretized frequency bins, over finite ranges of signal and idler frequencies. Poor choice of discretization and spectral range can give incorrect results. To quantify this, we characterize the effect of discretization and spectral range on inferred spectral separability.

What's more, in many situations only the \emph{joint spectral intensity} (JSI) can be measured directly. Many papers use the $\sqrt{\mathrm{JSI}}=|\mathrm{JSA}|$ to get information about the photons' spectral separability, but this neglects the effect of possible sign changes, or temporal correlation introduced by e.g. a chirped pump.  To quantify this, we also characterize the effect of discretization and spectral range on the inferred photons' spectral separability  when computed by taking the Schmidt decomposition of the $\sqrt{\mathrm{JSI}}$ and the JSI. 

\clearpage

\section{Spectral properties of the two-photon downconverted state} \label{sec:sec1}
We start by reviewing the spectral properties of downconverted photon pairs and their use for heralded single photon generation.

\subsection{Joint spectral amplitude}

The PDC process mediates the conversion of high-energy pump $p$ photons into pairs of lower energy photons, historically known as the signal $s$ and idler $i$ photon. 
Due to the spontaneous nature of photon-pair creation, the generated PDC state $\ket{\psi_{\mathrm{pdc}}}$ is described as a superposition of a large vacuum term, a term with a single photon-pair, and terms corresponding to higher-order photon-pair events. The single photon-term dominates when PDC is used for single photon generation in heralded or post-selection schemes. 
 In this paper, we therefore focus on the single-photon pair term and its spectral properties, neglecting higher photon numbers. 
The two-photon term of the full PDC state, assuming Type-II downconversion in a single-mode one dimensional propagation geometry, is \cite{grice1997spectral}
\begin{align}\label{eq:twophot}
\ket{\psi_{\mathrm{2}}}{}\propto{}&\iint \mathrm{d}\omega_{i}  \mathrm{d}\omega_{s}f(\omega_i,\omega_s)  \ket{\omega_{i}}_{i}\ket{\omega_{s}}_{s}\,,
\end{align}
where $\ket{\omega_{i}}_{j}$ is a one-photon Fock state of frequency $\omega_{i}$ prepared in mode $j$.  We take  $f(\omega_{i},\omega_{s})$  to be normalized such that $\int d\omega_{i}d\omega_{s}|f(\omega_{i},\omega_{s})|^2 =1$. We consider collinear single-mode PDC emission (state-of-the-art experimentally) and we neglect transversal and multi-mode effects.

PDC photon pairs are characterized by the pump envelope function (PEF) $\alpha(\omega_p)$ (where $\omega_p=\omega_i+\omega_s$ due to energy conservation), and the material properties of the crystal, which are captured by the phase-matching function (PMF) $\phi(\omega_{i},\omega_{s})$:
\begin{equation}\label{eq:JSA}
f\ooa = \alpha (\omega_i+\omega_s)  \phi\ooa \, .
\end{equation}
The PMF accounts for the crystal's dispersion, as well as longitudinal variations in the crystal's nonlinearity. We define the PMF as:
\begin{equation}
\phi(\omega_{i},\omega_{s}) = \int_{0}^{L} g(z) \ e^{i \Delta k (\omega_{i},\omega_{s} ) z} dz\,,
\label{generalPMF}
\end{equation}
where $L$ is the length of the crystal, $\Delta k (\omega_{i},\omega_{s})= k_p  (\omega_{i}+ \omega_{s}) - k_i (\omega_{i})- k_s (\omega_{s})$ where $k_j(\omega)=\omega n_j(\omega)/c$, and $g(z) = \chi^{(2)}(z)/\chi^{(2)}_0$ is the normalised nonlinearity along the crystal. We take $g(z)$ to really mean $g(z)\rightarrow g(z)\Pi_{0,L}(z)$, where $\Pi_{0,L}(z)$ is a rectangular function $\Pi_{0,L}(z)=1$ for $0<z<L$ and $\Pi_{0,L}(z)=0$ otherwise.

To simplify the discussion, we expand the wave numbers to first order $k_j(\omega)=k_j(\bar\omega_j)+v_j^{-1}\Omega_j$, where $v_j = d \omega / d k_j(\omega)|_{\omega=\bar\omega_j}$ is the group velocity of photon $j$, $\Omega_j=\omega_j-\bar\omega_j$, and $\bar\omega_p=\bar\omega_s+\bar\omega_i$. We can ignore quadratic and higher-order terms corresponding to group velocity dispersion if  the photons in each mode are not too spread out around the central frequencies. We can then write
\begin{align}
\Delta k = \Delta k_0+(v_p^{-1}-v_i^{-1})\Omega_i +(v_p^{-1}-v_s^{-1})\Omega_s \,,
\end{align}
where $\Delta k_0= k_p(\bar\omega_s+\bar\omega_i)-k_s(\bar\omega_s)- k_i(\bar\omega_i)$.

\begin{figure}[t]\center
\includegraphics[width=0.49\textwidth]{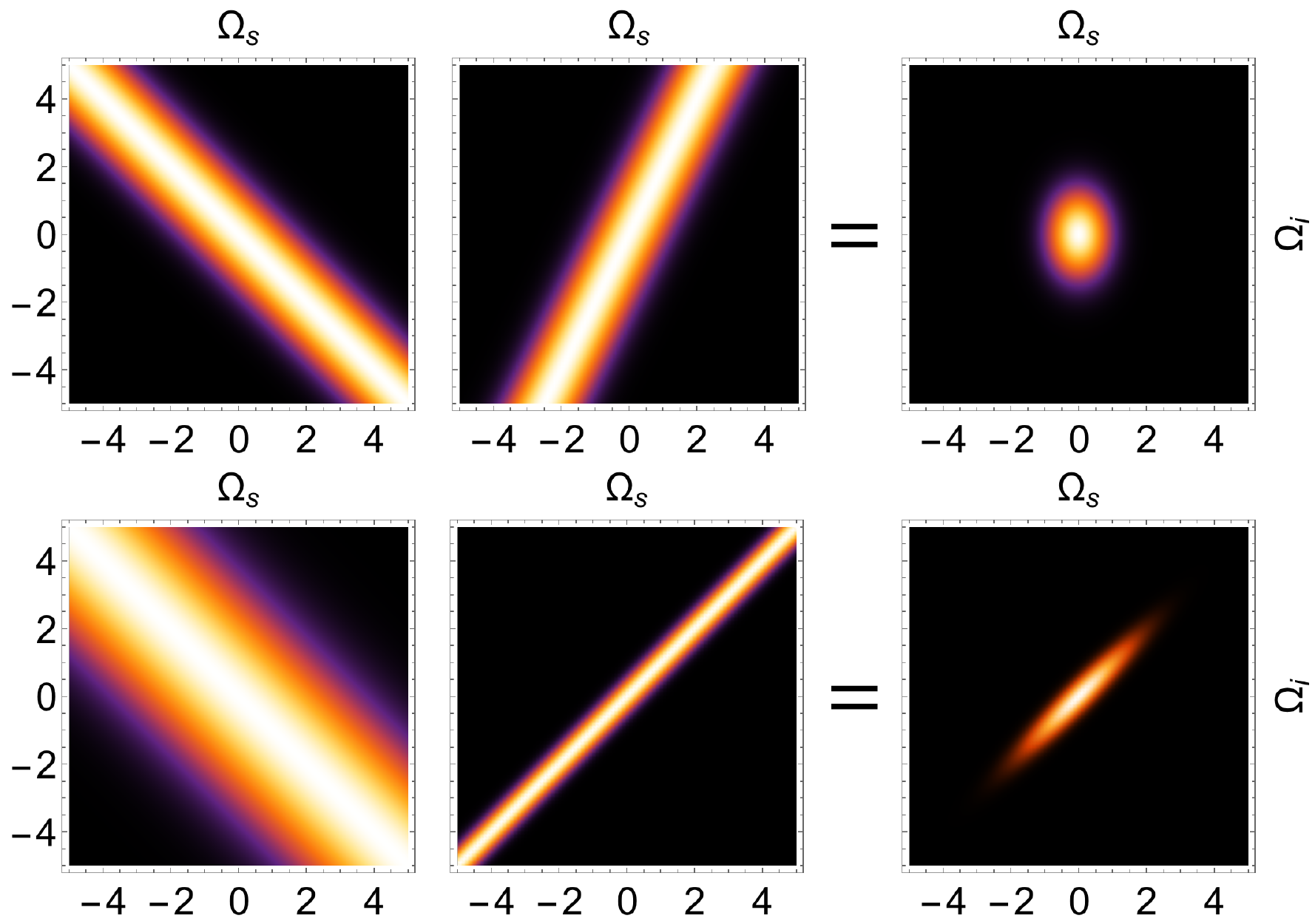}
\caption{Example of JSAs (right) composed from Gaussian PEFs (left) and Gaussian PMFs (middle). The top-right JSA is perfectly separable, while the bottom-right JSA is highly correlated.
}
\label{fig:JSA_general_case}
\end{figure}

Two example JSAs (one perfectly separable, the other highly correlated) composed from Gaussian pump and phase-matching functions, with $\Delta k_0=0$, are shown in Fig. \ref{fig:JSA_general_case}. When plotted  as a function of $\Omega_i$ and $\Omega_s$, the  PEF is always oriented along the anti-diagonal, while the PMF lies along an axis defined by the angle $\theta$, which depends on the group velocities according to
 \begin{align}\label{eq:theta}
\tan \theta = -\frac{v_p^{-1} - v_s^{-1}}{v_p^{-1}- v_i^{-1}}\,.
\end{align}
Picking group velocities appropriately is known as \emph{group velocity matching} (GVM). 

The JSA completely characterizes the spectral properties of a two-photon downconverted state, and will be the focus of this paper. 

\subsection{Schimdt decomposition}

To simplify calculations, the JSA can be expressed as a sum of orthogonal modes:
\begin{align}
  f\ooa = \sum_k b_k q_k(\oos) r_k(\ooi)\,,
  \end{align}
in what is known as the \emph{Schmidt decomposition} \cite{law2000continuous,laudenbach2016modelling}. The \emph{Schmidt coefficients} $\{b_k\}$ are real numbers that sum to unity if $f\ooa$ is normalized, and the \emph{Schmidt modes} $\{q_k(\oos)\}$ and $\{r_k(\ooi)\}$ are orthonormal single-photon spectral functions.

Likewise, the two-photon state (\ref{eq:twophot})  can be decomposed as
\begin{align}\label{eq:decomposedpdc}
\ket{\psi_{\mathrm{2}}}{}={}& \sum_k b_k \ket{q_k }_s  \ket{r_k}_i \, ,
\end{align}
where
\begin{align}
\ket{q_k }_s={}&\int  d\omega q_k (\omega)  \ket{\omega}_{s};~~~ \ket{r_k}_i={}\int d\omega r_k(\omega)   \ket{\omega}_{i}\,,
\end{align}
 are orthonormal states in the signal and idler subspaces. The states satisfy the orthonormality conditions $\bra{q_k }{q_{k'} }\rangle_s=\delta_{kk'}$ and $\bra{r_k }{r_{k'} }\rangle_i=\delta_{kk'}$, which simplifies expressions for many interesting quantities that can be written just in terms of the Schmidt coefficients. We will see an example of this in the next section.

\subsection{Spectrally Pure Heralded Photons}\label{sec:pur}

A drawback of PDC is that photon pairs are generated spontaneously, making them difficult to interfere in optical networks. The spontaneous nature of the downconverted source can be mitigated by placing a photon detector in one of the downconverted modes. Photon-number correlations between the two downconverted modes ensure that detection of a single photon in one mode projects the state in the other mode into a single photon---a process known as \emph{heralding}. The heralded photon can then be stored for future use in a quantum memory or appropriately delayed so that it arrives in the experiment at the right time \cite{kaneda2017quantum}.

To calculate the heralded  state, in say mode $s$, we model single-photon detection with a flat frequency response, in say mode $i$, by the projector 
\begin{align}
\hat{P}_{i}=\int d\omega\ket{\omega}_{i}\bra{\omega}_{i}= \sum_k  \ket{r_k}_i\bra{r_k}_i\,,
\end{align}
expressed in terms of the Schmidt modes $\ket{r_k}_i$ for convenience. 

The heralded state is then calculated by applying the Born rule, and tracing out the detected mode:
\begin{align}
\rho_{s}={}&\mathrm{Tr}_{i}\left[\ket{\psi_{\mathrm{pdc}}}{}\bra{\psi_{\mathrm{pdc}}}{}(\hat{\mathbb{I}}_s\otimes\hat{P}_{i})\right]\\
={}& \sum_k b^2_k \ket{q_k }_s  \bra{q_k }_s \,.
\end{align}
This result shows that, after detection of a single photon in mode $i$, the state in mode $s$  is a statistical mixture of single-photon states with orthogonal spectral distributions $q_k(\omega)$. The mixed nature of this state is undesirable because it reduces its interference visibility in an interferometric network \cite{branczyk2017hong}. 

The degree to which the state is mixed can be quantified by the purity: 
\begin{align}\label{eq:purity}
P_s = \text{Tr}\left[ \rho_s^2 \right] = \sum_k b_k^4  \,,
\end{align}
which ranges from $P_s=1$ for a pure state to $P_s=1/N$  (where $N$ is the number of Schmidt modes) for a maximally mixed state. 

When $b_0=1$ and all other coefficients are zero, $\rho_{s}= \ket{q_0 }_s  \bra{q_0 }_s$ is a pure state, and $P_s=1$. The JSA that leads to this is a separable JSA: $f\ooa =  q_0(\oos) r_0(\ooi)$. To achieve high-purity heralded single photons, the JSA must be separable.

\subsection{JSA Separability in the ideal case}

It was well-known that, under certain conditions, Gaussian pump and phase-matching functions can make the JSA separable \cite{u2006generation}. Recently, it was shown that Gaussian functions are the \emph{only} functions that make the JSA separable \cite{quesada2018gaussian}. In this section, we briefly review the conditions for perfect separability. A detailed study of these conditions can be found in \cite{u2006generation}.

We define a Gaussian pump function as 
\begin{align}
&\label{eq:gaussPump} \alpha_{\text{Gauss}} \ooa = \exp\left[-\frac{\left( \Omega_s +\Omega_i   \right)^2}{2 \sigma_{\textsc{pef}}^2}\right]  \,.
\end{align}
We also define the \emph{bandwidth} (or simply \emph{width}) of a spectral (or temporal) distribution as its FWHM: $2 \sqrt{2 \log{2}}\ \sigma_{\textsc{pef}} $ for a Gaussian PEF. We stress that when defining the bandwidth, we are working with  the spectral \emph{amplitude} of the pump, while in other studies the bandwidth was defined differently (either with reference to the profile of the spectral \emph{intensity} of the pump, or by considering the 1/e width instead of the FWHM  \cite{u2006generation,branczyk2010optimized,branczyk2011engineered,weston2016efficient}).
 
We define a Gaussian PMF as
\begin{align}
&\label{eq:gaussSymPMF} \phi_{\text{Gauss}} \ooa = \exp\left[-\frac{\left( \sin(\theta)\Omega_s -\cos(\theta)\Omega_i \right)^2}{\sigma_{\textsc{pmf}}^2}\right]  \, ,
\end{align}
where  $\theta$ defines the orientation of the PMF and depends on the group velocities according to Eq. (\ref{eq:theta}) \cite{jin2013widely,kaneda2016heralded,laudenbach2017numerical}. 

Perfect separability happens when $0<\theta< \pi/2$ and $2\cos(\theta)\sin(\theta)= \sigma_{\textsc{pmf}}^2/ \sigma_{\textsc{pef}}^2$. The examples in Fig \ref{fig:JSA_general_case} correspond to $\theta\approx\pi/3$ (top) and $\theta=\pi/4$ (bottom). In the example on top, the photons have different bandwidths and won't interfere well with each other, but they are suitable in a heralded configuration because the JSA is serparable. At the bottom, the photons will be indistinguishable (having the same spectral bandwidth and shape) but they are not spectrally pure because the JSA is correlated.

For the remainder of the paper, we will focus on the special case where  $v_p^{-1} = \left(v_s^{-1} + v_i^{-1}\right)/2$, i.e. $\theta=\pi/4$. In this case, the PMF is perpendicular to the PEF, and perfect separability happens when $\sigma_{\textsc{pmf}}^2= \sigma_{\textsc{pef}}^2$, such that:
\begin{align}
&\label{eq:gaussSymPMF} \phi^{\text{sym}}_{\text{Gauss}} \ooa = \exp\left[-\frac{\left(\Omega_s -\Omega_i \right)^2}{2\sigma_{\textsc{pmf}}^2}\right]  \, .
\end{align}
This regime is known as the symmetric GVM condition, which, for Gaussian functions, generates separable photons with equal bandwidths. These photons can be used for heralded photon generation, but the photons can also both be fed into an experiment, as they will exhibit perfect two-photon interference.

\section{Experiment design}\label{sec:exp} \label{sec:sec2}
In this section, we study how realistic pump fields and realistic nonlinear crystals with fabrication imperfections impact spectral separability.

\subsection{Typical pump and phase-matching functions}

Many studies of the joint spectral properties of downconversion pairs assume either \textit{Gaussian} or \textit{delta} PEFs \cite{grice1997spectral,keller1997theory,erdmann2000restoring,grice2001eliminating,kim2002generation,giovannetti2002extended,giovannetti2002generating,u2006generation,christ2009pure,kaneda2016heralded,weston2016efficient,laudenbach2017numerical,barbieri2017hong}. These functions are convenient to work with analytically, but often don't reflect what happens in an experiment. Most experiments which aim to create more than just one heralded photon, or multiple photon pairs, are performed with mode-locked, ultra-short-pulsed lasers whose temporal intensity function can be described by a squared hyperbolic secant (sech) function, $\mathrm{sech}^2 (t/\tau)$, where $\tau$ is a temporal scaling factor. This yields a PEF represented, up to an irrelevant linear phase, by a sech function
\begin{align}
&\label{eq:sechPump} \alpha_{\text{sech}} \ooa = \text{sech} \left[ \frac{1}{2} \pi \tau \left( \Omega_s +\Omega_i  \right)  \right] \, ,
\end{align}
with a bandwidth of $4 \cosh^{-1}{[2]}/(\pi \tau)$, which we define as the FWHM of the PEF. The sech and Gaussian PEFs  have equal width when $\tau \approx 0.712\sigma_{\textsc{pef}}$.

\begin{figure}[h]\center
\includegraphics[width=0.5\textwidth]{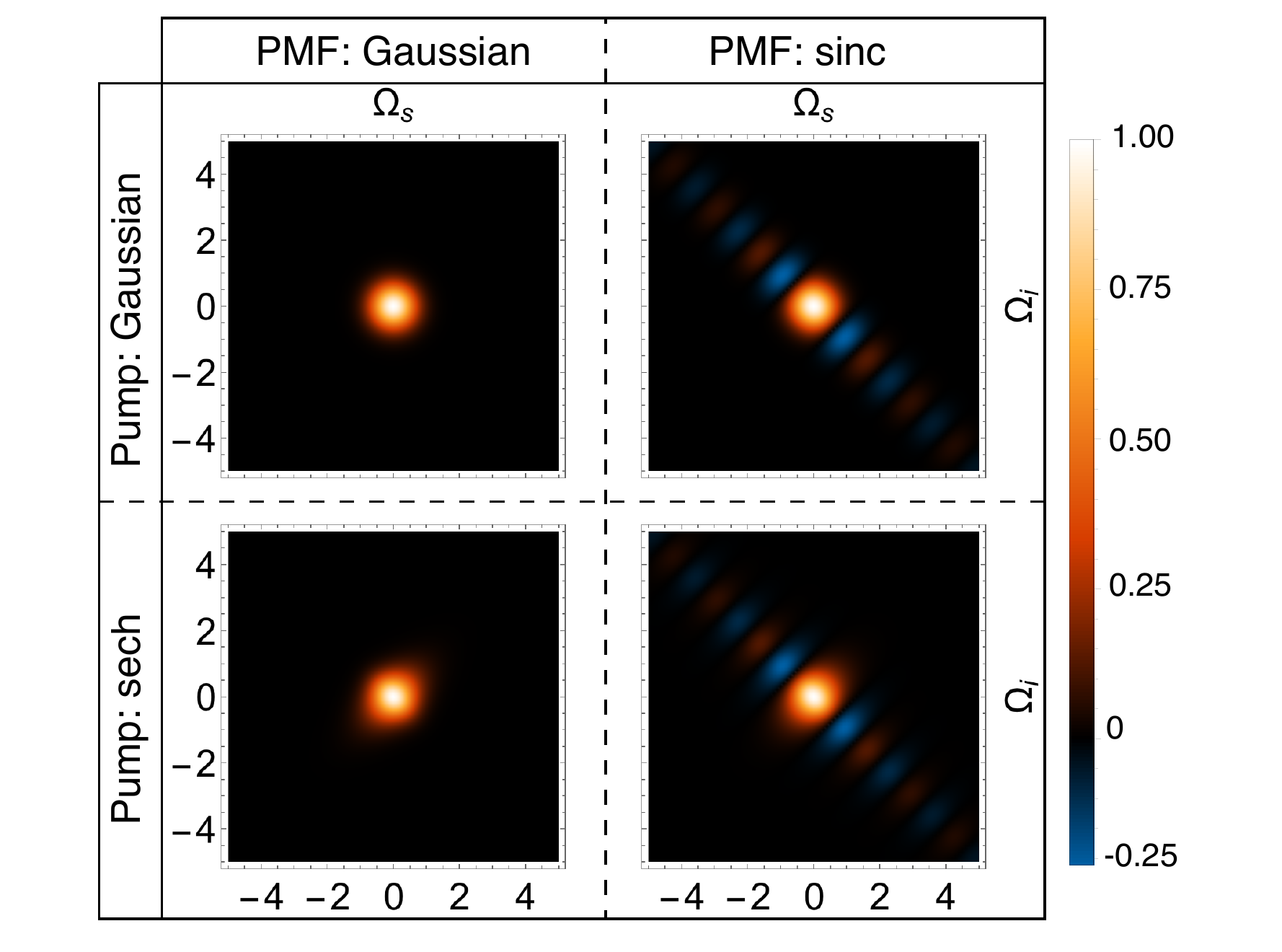}
\caption{Joint spectral amplitudes in symmetric group velocity matching condition for four different combinations of pump functions and phase-matching function.}
\label{fig:JSAs}
\end{figure}

We also consider realistic PMFs.  Most crystals either have a constant nonlinearity profile or are periodically poled. In both cases, this leads to a sinc-shaped phase-matching function. We define a sinc-shaped PMF oriented at $\theta = \SI{45}{\degree}$ as
\begin{align}
&\label{eq:sincSymPMF} \phi^{\text{sym}}_{\text{sinc}} \ooa = \text{sinc}\left[ \kappa \left(  \Omega_s - \Omega_i) \right) \right] \, .
\end{align}

The sinc and Gaussian PMFs  have equal width when $\sigma_{\textsc{pmf}} \approx 1.61/\kappa$. Also, the sech PEF and sinc PMF have equal width when  $\tau \approx 0.442\kappa$.

\begin{figure}[t]\center
\includegraphics[width=0.45\textwidth]{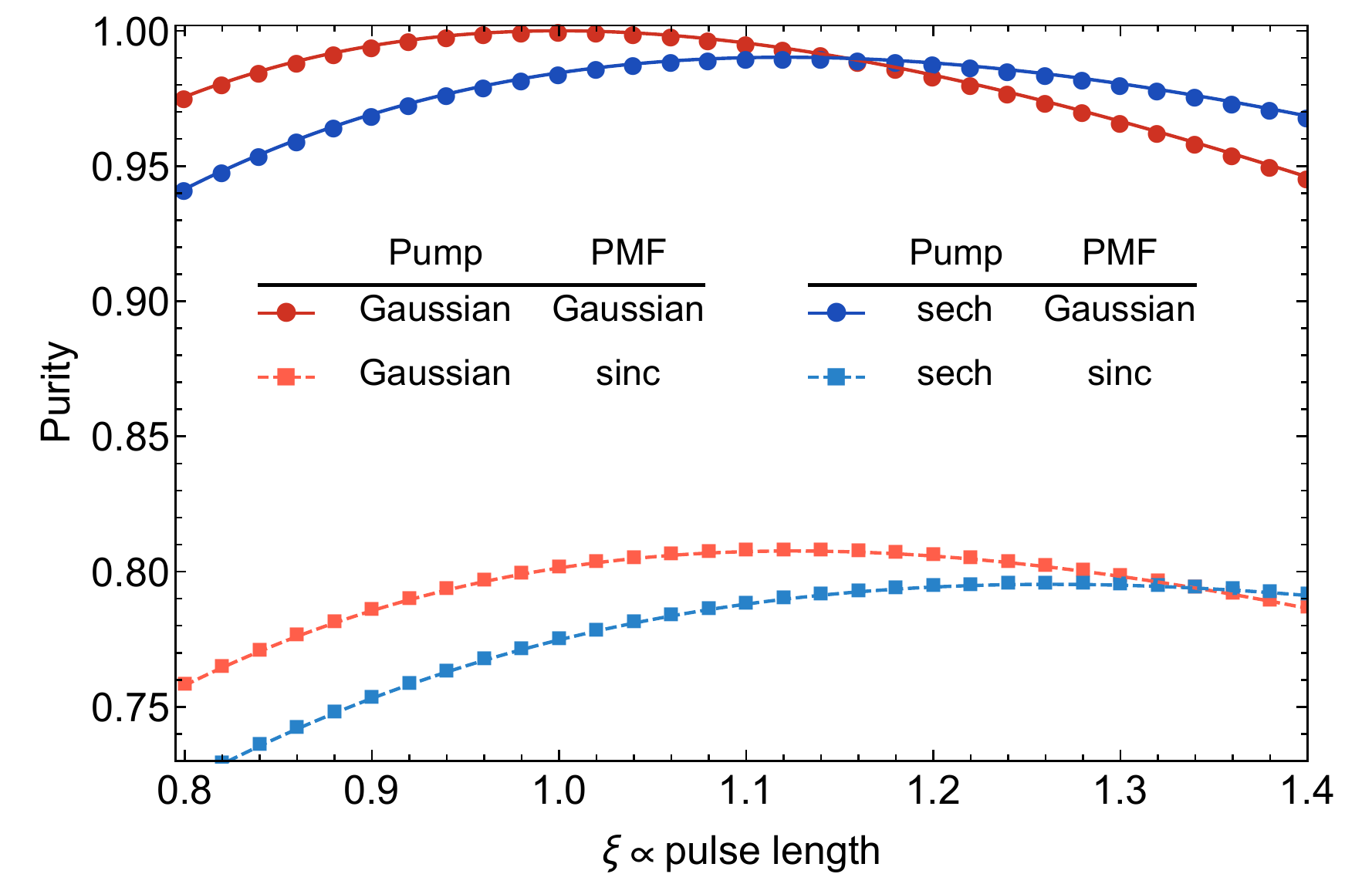}
\caption{Heralded single-photon spectral purity at different pump widths for four combinations of PEF and PMF shapes in symmetric GVM condition. The parameter $\xi$ is the ratio between the widths of the PMF and PEFs, defined in Eq. \ref{eq:xi}.
}
\label{fig:pulse_duration}
\end{figure}

We now analyze how realistic PEFs and PMFs affect JSA separability. The JSAs given by the four aforementioned PEF and PMF combinations are shown in Fig \ref{fig:JSAs}. For all four combinations, the spectral purity of heralded photons depends on the relative widths of the functions. To maximize the purity, we define the parameter $\xi$ as the ratio between the widths of the PMFs and PEFs: 
\begin{align}\label{eq:xi}
\xi=\frac{\sigma_{\textsc{pmf}}}{\sigma_{\textsc{pef}}}\approx1.40\ \sigma_{\textsc{pmf}}\tau \approx \frac{1.61}{\sigma_{\textsc{pef}}\kappa} \approx 2.26 \frac{\tau}{\kappa }\, ,
\end{align}
and optimize over $\xi$.  As is well known,  the optimal ratio for a Gaussian-Gaussian combination is $\xi=1$, but we show that for other combinations, this can vary by up to  \SI{26}{\percent}. A sech PEF reduces the maximum purity only slightly, while a sinc PMF reduces the maximum purity significantly. Table \ref{tab:pxi} shows maximum purities and corresponding $\xi$ for all four PEF-PMF combinations.  
\begin{table}[h!]
\begin{tabular}{cccc} PEF & PMF & maximum $P_s$ & optimal $\xi$ \\\hline Gaussian & Gaussian & 1 & 1  \\ sech & Gaussian & 0.99 & 1.12\\ Gaussian & sinc & 0.80 & 1.13 \\ sech & sinc & 0.79 & 1.26 \\ \end{tabular}
\caption{Maximum purities and corresponding $\xi$ for the most common combinations of pump envelope functions and phase-matching functions.}
\label{tab:pxi}
\end{table}
\\
Fig. \ref{fig:pulse_duration} shows the dependence of the purity on $\xi$ for all four PEF-PMF combinations.

Clearly, the experimentally very common practice of establishing a FWHM pulse length, or a spectral bandwidth, of an ultrafast pump laser, and then simply converting it to a supposedly equivalent Gaussian PEF can lead to drastic mis-calculations of required PMF bandwidths and thus crystal lengths. The correct procedure would be to e.g. use an auto-correlator and deconvolve the temporal auto-correlation trace into the function which most accurately represents the actual PEF shape. The resulting PEF should be compared to an accurate spectral bandwidth measurement to make sure the pulse is transform limited (more on that in section \ref{subsec:chirp}). In some cases, the measured PEF might not be represented accurately by either one of the analytical functions, in which case the optimal relation between PMF and PEF widths should be determined numerically \cite{quesada2018gaussian}.

\begin{figure}[b]\center
\includegraphics[width=0.45\textwidth]{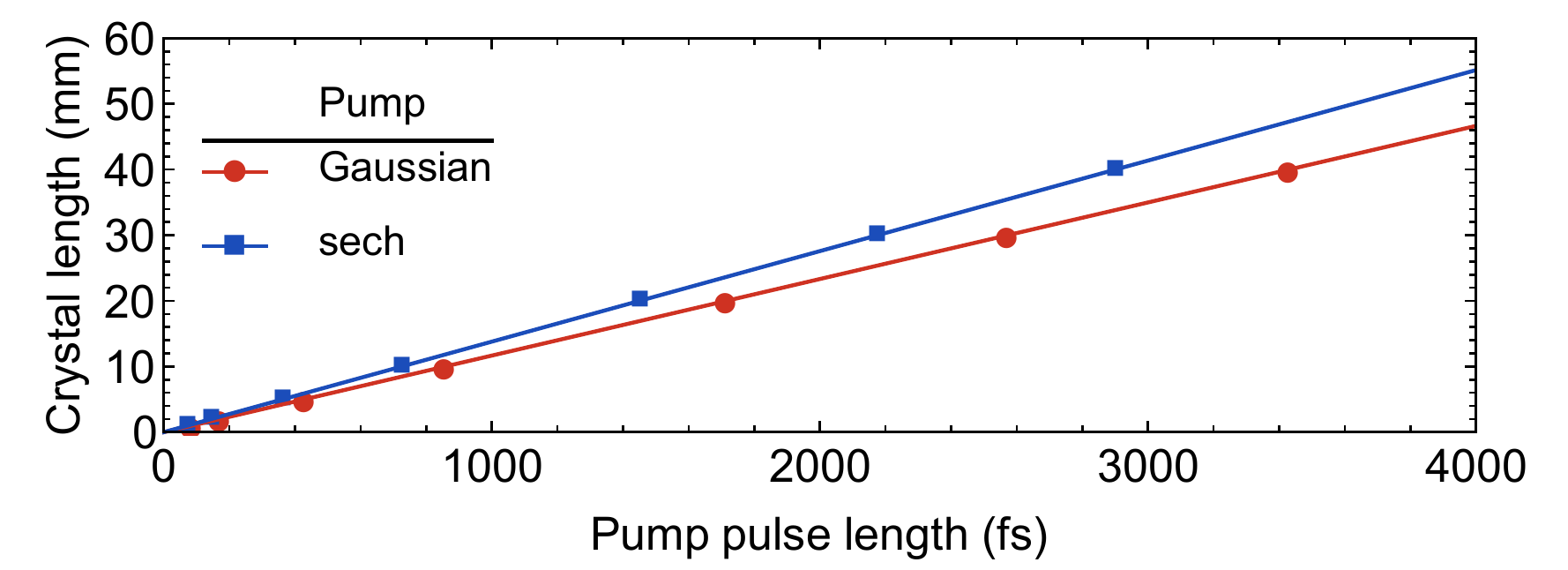}
\caption{Crystal length as a function of  pump pulse duration for periodically poled KTP. Note that the pump duration is defined as the FWHM of the temporal intensity profile of the pulse (that can be measured, e.g., via an autocorrelation measurement). The $\Delta k$ dependence on the pump, signal and idler frequencies is computed from  the Sellmeier equations in \cite{fradkin1999tunable,emanueli2003temperature,konig2004extended}.}
\label{fig:crystal_length}
\end{figure}

For unpoled or periodically poled crystals, the crystal length determines the PMF width (this is not the case for custom-poled crystals, which will be discussed later). When designing an experiment, one chooses the crystal length based on the PEF  FWHM (or vice-versa). But the exact form of the relationship between the optimal crystal length and PEF FWHM differs for different PEF shapes. As a concrete example, we consider a periodically-poled KTP crystal, pumped  with either Gaussian or sech PEFs centred at \SI{791}{\nano\metre}, in the  symmetric GVM regime. Fig. \ref{fig:crystal_length}, shows that the crystal length that optimizes JSA separability is different for Gaussian or sech PEFs.

The spectral correlations that arise from non-Gaussian pump and phase-matching functions can be filtered out. This, however, comes at a price. Spectral filtering acts at the intensity level and can destroy photon-number correlations between the two downconverted modes \cite{branczyk2010optimized,Laiho2011,Christ2014a,meyer2017limits}. This reduces the photon-number purity of the heralded state. It also spoils interference between the downconverted modes when  combined in an optical network. 

Ideally, the spectrum  could be shaped at the amplitude level. The PEF amplitude can be shaped using optical pulse shaping \cite{Weiner2011ultrafast}. The PMF amplitude can be shaped using nonlinearity shaping methods, which will be discussed in  Section \ref{sec:nls}. A combination of optical pulse shaping and nonlinearity shaping can reduce correlations in the JSA   \emph{without} destroying photon-number correlations.

\subsection{Chirped pump functions} \label{subsec:chirp} 
Establishing group-velocity matching conditions via JSA simulations and then joint spectral measurements is now common practice. It's important to note though that spectral measurements of e.g. the pump pulse do not contain any information about the temporal pulse duration. 

It's sometimes been claimed that high-quality two-photon interference cannot be achieved between photons from independent sources pumped by temporally `long'  pulses with durations of picoseconds and above---supposedly because this increases uncertainty in when the photons leave the crystal, thus increasing uncertainty in when they meet at the beamsplitter. But this interpretation is wrong because perfect two-photon interference can always be achieved as long as the heralded photons are pure, which happens if the PEF and PMF bandwidths are matched, regardless of the pump's temporal length (this is the key principle of the GVM technique for PDC sources). 

Other timing uncertainties, however, can still reduce two-photon interference, e.g. two pulsed lasers pumping separate heralded photon sources might drift out of sync \cite{kaltenbaek2006experimental}. Another common scenario is that a laser pulse might not be transform-limited, i.e. the temporal duration might exceed the time-bandwidth product. This is the case we will study here.

So far, we have considered Fourier transform-limited PEFs. However, when short optical pulses propagate through a transparent medium whose refractive index is wavelength dependent, they acquire a phase that depends nonlinearly on the wavelength, known as frequency \emph{chirp}.
To study how a \emph{linear} frequency chirp affects the downconverted photons, we introduce  a \emph{quadratic} spectral phase to the PEF, i.e., we multiply the PEF by $e^{-i k \left(\bar\omega_p-\omega_p\right)^2}$ (or, equivalently, multiply the JSA by $e^{-i k \left( \Omega_s + \Omega_i \right)^2}$), where $k$ is equal to half of the group delay dispersion in the material \cite{rulliere2005femtosecond}. This phase delays the pulse and introduces temporal broadening, which introduces phase correlations in the JSA, reducing the  spectral purity of heralded photons. 

\begin{figure}[t]\center
\includegraphics[width=0.45\textwidth]{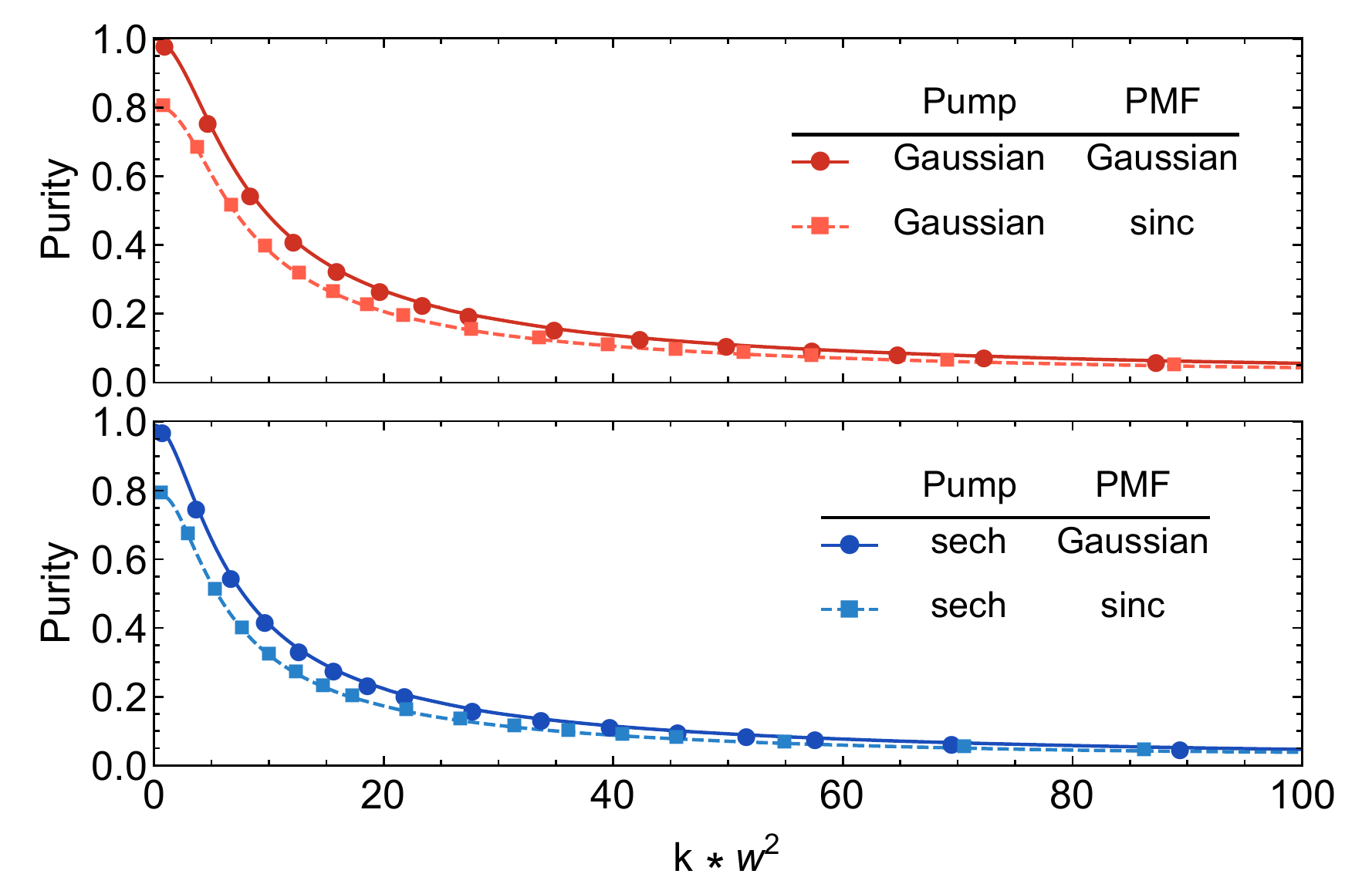}
\caption{
Heralded single-photon spectral purity at different $k w^2$ values for four combinations of PEF and PMF shapes in symmetric GVM condition. When $k w^2 = 0$, the pulse is transform-limited, while large values of $k w^2$ correspond to a strongly chirped pulse.
}
\label{fig:chirped_purity}
\end{figure}

The spectral purity of a chirped JSA can be parametrized by the dimensionless parameter $k  w^2$, where $w$ is the spectral width of the PEF. This tells us that there is a trade-off between the  pump width and the amount of chirp that can be tolerated, i.e.,  increasing the chirp or increasing the square of the width of the pump will have the same effect on the purity. Our numerical simulations show that the purity decays almost exponentially as $k  w^2$ increases. This can be seen in  Fig. \ref{fig:chirped_purity}, where we plot the heralded-photon purity as a function of $k  w^2$.

\begin{figure}[b]\center
\includegraphics[width=0.45\textwidth]{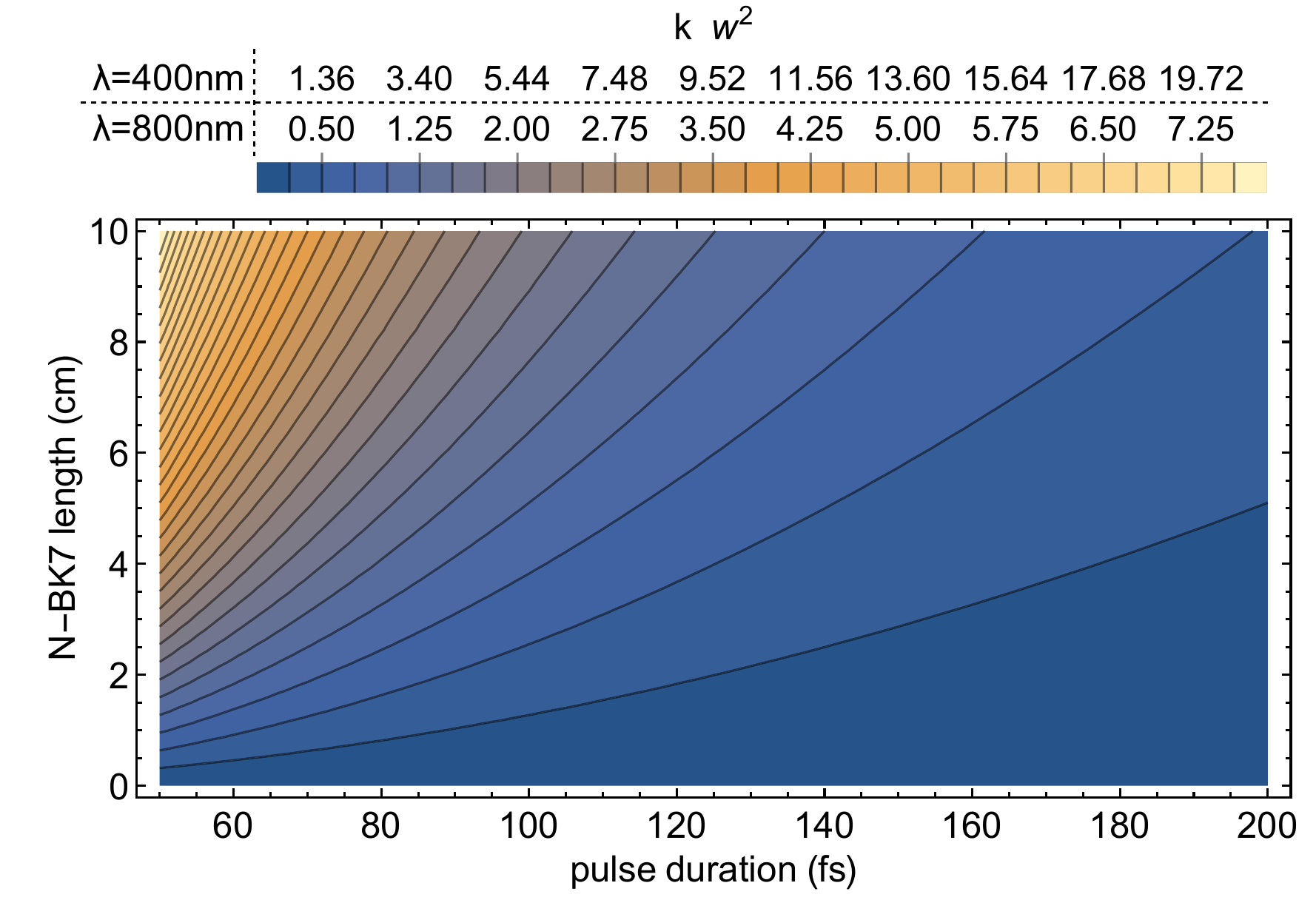}
\caption{
$k w^2$ in N-BK7 for a sech pulse at different pulse duration and central wavelengths.
}
\label{fig:GDD_NBK7}
\end{figure}

To explore the trade-off between $k$ and $w^2$, we model sech pulses propagating in optical glass N-BK7. For example, a \SI{400}{\nano\meter}, \SI{50}{\femto\second} sech pulse (where we define the pulse length as the FWHM of the temporal \textit{intensity} profile), which is the pulse length of modern GHz repetition rate Ti:sapph lasers, passing through \SI{1}{\centi\meter} of the standard optical substrate of N-BK7 acquires a quadratic phase of $k w^2 \approx 2.1$, decreasing the purity from $\sim0.99$ to $\sim0.90$ (Gaussian PMF) or from $\sim0.80$ to $\sim0.74$ (sinc PMF). But a \SI{200}{\femto\second} sech pulse travelling through the same piece of glass acquires a quadratic phase of only $k w^2 \approx 0.13$, decreasing the purity by  less than \SI{0.1}{\percent}. However, if the same \SI{200}{\femto\second} pulse is sent through \SI{30}{\centi\meter} of fused-silica single-mode fibre for spatial mode filtering, the corresponding chirping is $k w^2 \approx 3.2$ and is therefore not negligible.
This shows that while group delay dispersion, and therefore spectral chirping, can be neglected for relatively long pulses (in particular picosecond pulses), it should be taken into account for short (i.e. sub-picosecond) pulses. Fig. \ref{fig:GDD_NBK7} shows values of $k w^2$ in N-BK7 for a sech pulse at different pulse durations and central wavelengths. 

While the frequency chirp introduces correlations in the JSA, these correlations \emph{are not} visible in the JSI. Therefore, in the presence of chirp, the $\sqrt{\mathrm{JSI}}$ is not a good indicator of heralded photon spectral purity, as will be discussed in Section \ref{sec:JSAJSI}.

\subsection{Nonlinearity shaping}\label{sec:nls}

In previous sections, we saw that although a Gaussian PMF is optimal for JSA separability, most standard crystals have sinc PMFs.  In this section, we discuss how Gaussian-shaped PMFs can be achieved through inversion of the crystal lattice at appropriate positions within the material, using e.g. ferroelectric \emph{poling}. 

The most common type of poling inverts the crystal lattice periodically to induce \emph{quasi-phase-matching} \cite{fejer1992quasi}. This doesn't change the shape of the PMF, but shifts its peak in  $\Delta k$ space to allow photon generation at desired frequencies. 

\begin{figure}[h]\center
\includegraphics[width=0.453\textwidth]{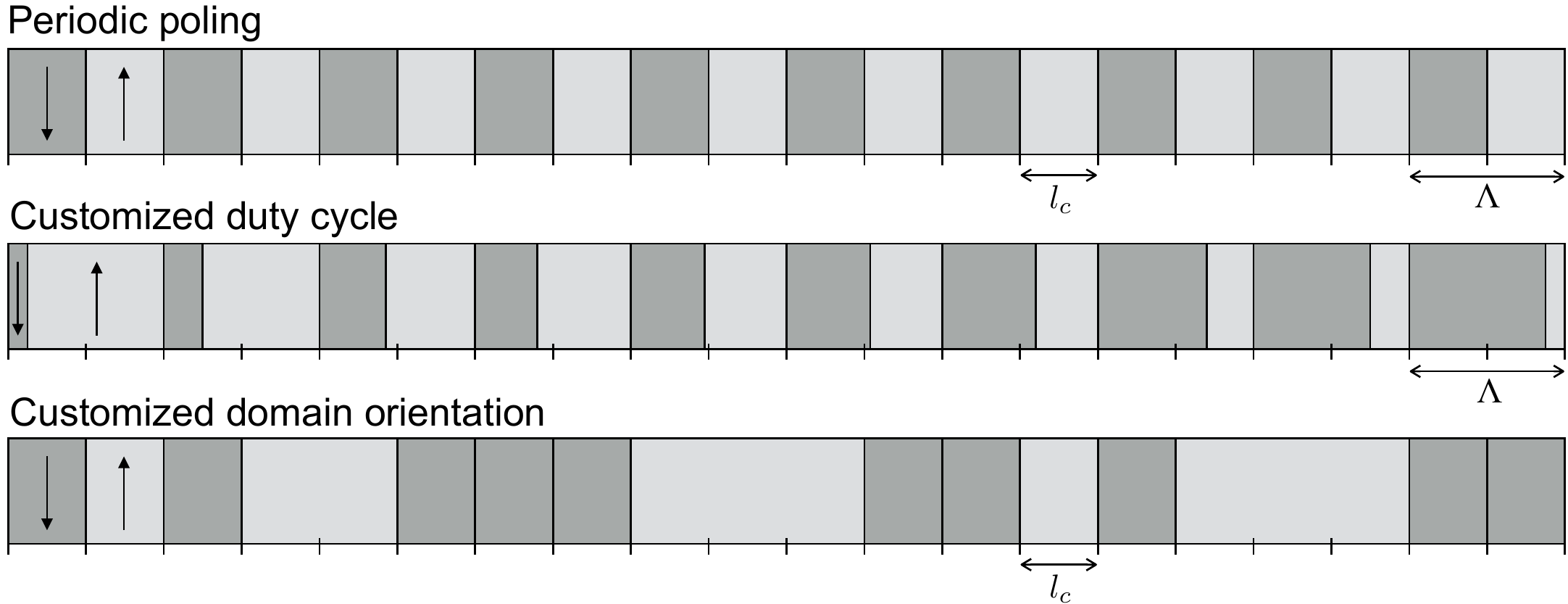}
\caption{Examples of three poling methods. Periodic poling: periodically alternating fixed-width domains shift the PMF. Customized duty cyle: periodically alternating domains with a customized duty cycle also shift and shape the PMF. Customized domain orientation: fixed-width domains with customized orientations shift and shape the PMF. }
\label{fig:pm}
\end{figure}

PMF shaping requires more complex poling patterns. 
One approach introduces domain-width variation to a pre-defined poling pattern. The most common case is customizing the duty-cycle of a periodically poled structure \cite{dixon2013spectral,chen2017efficient} .
This is a well-know technique adapted from nonlinear optics application in the classical regime: however, it doesn't yield near-unity purity \cite{graffitti2017pure}.
Another approach introduces aperiodicity to the poling, while keeping the width of the poled domains fixed \cite{branczyk2011engineered,dosseva2016shaping,tambasco2016domain,graffitti2017pure}: we refer to these methods collectively as \emph{customized domain-orientation} methods.
While in the long crystal limit any of these methods provides nearly-separable JSAs, for short crystal matched with femtosecond lasers, deterministic sub-coherence length domain engineering is required to achieve spectrally pure heralded-photons \cite{graffitti2017pure}.
Customization of domain widths of pre-customized poling patterns has also been proposed  \cite{graffitti2017pure} and experimentally implemented \cite{graffitti2018independent}, demonstrating high visibility in the interference of unfiltered heralded-photons on a BS---an optimal benchmark for the PDC spectral purity, as shown in section \ref{sec:HOM}.
Fig \ref{fig:pm} shows a schematic representation of some of these poling methods. 

For simplicity, we will compare the customized duty-cycle method proposed in \cite{dixon2013spectral} with the customized domain-orientation method proposed in \cite{tambasco2016domain}.

\subsubsection{Modelling customized poling structures}

To model the PMF of a customized structure, we consider a crystal divided into $N$ domains. The shape of the PMF $\phi(\Delta k )$ for such a crystal arises from interference between the PMFs for individual domains, $\phi_j(\Delta k)$, with relative phase-shifts $s_j=\pm1$ introduced by the domains' relative orientations. Each domain is just a short crystal with a constant nonlinearity $g(z)=1$, and therefore has a sinc-shaped PMF, with a phase determined by the domain's position. The PMF for the crystal is:
\begin{align}
 \phi(\Delta k ) ={}&\sum_{j=1}^{N}s_{j} \phi_j(\Delta k)\\\label{eq:phij}
 ={}& \sum_{j=1}^{N}s_j w_{j}\mathrm{sinc}\left(\frac{\Delta k~w_{j}}{2}\right)e^{i\Delta k z_{j}}\,,
\end{align}
where $w_{j}$ is the width of the $j$th domain centred at position $z_j$. We consider structures with adjacent domains, and for physical reasons, assume that the domains do not overlap. From Eq. (\ref{eq:phij}), we see  that $\phi(\Delta k)$  can be shaped by customizing the domains' relative orientations $s_j$, widths $w_j$ and central positions $z_j$.

\subsubsection{Fabrication imperfections in nonlinearity shaping using custom poling}

\begin{figure}[b]
\includegraphics[width=\columnwidth]{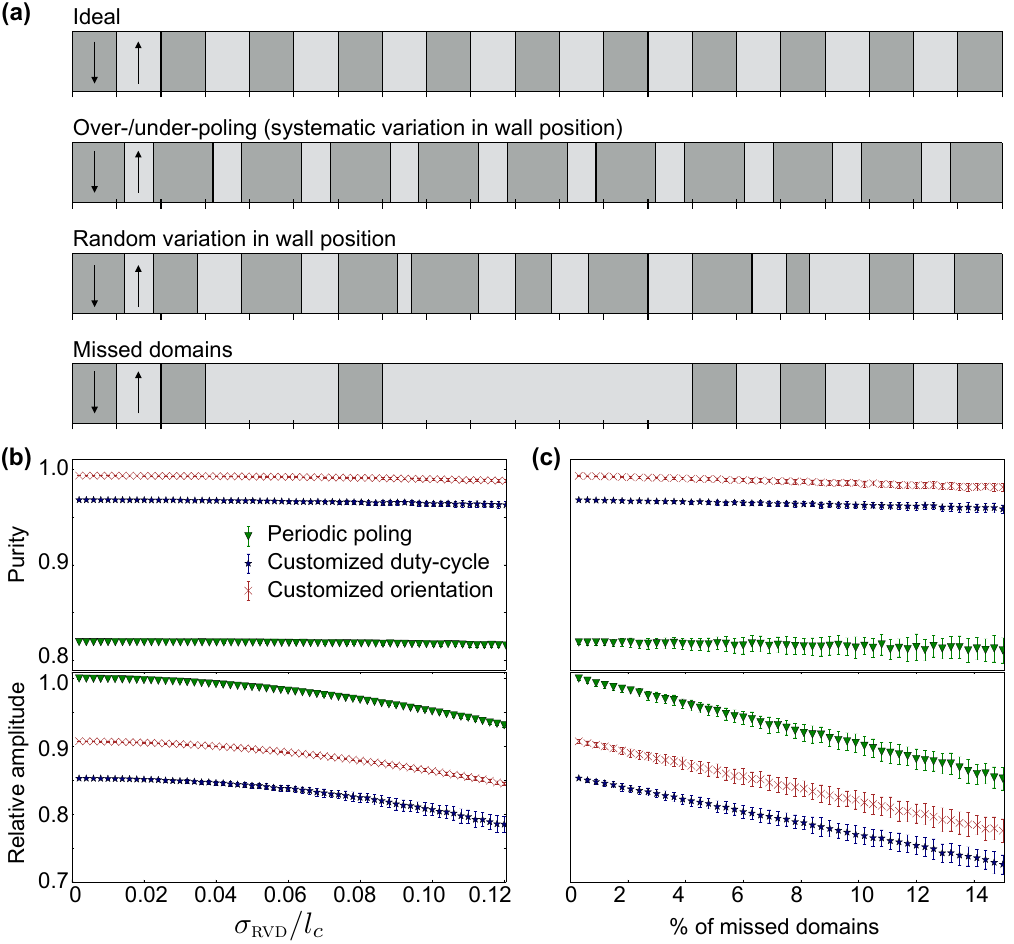}
\caption{(a) Examples of imperfection profiles on a periodically poled structure. (b) Spectral purity and relative amplitude with random variations in wall positions based on a Gaussian distribution centered around the expected position of each domain, parameterized by the standard deviation $\sigma_{\textsc{rvd}}$. (c) Spectral purity and relative amplitude with randomly chosen missed domains. }
\label{fig:fab}
\end{figure}

A popular method for generating poled crystals is ferroelectric poling, in which the spontaneous polarization of a ferroelectric crystal can be reversed under the influence of a sufficiently large electric field that is applied using lithographically defined periodic electrodes  \cite{Miller1998}. This process is susceptible to various fabrication imperfections:  timing errors in applying the  field may systematically over- or under-pole  inverted domains, roughness in electrode lithography may introduce random variations in  domain walls, and failure of the crystal to nucleate may prevent inversion, resulting in missed domains.  These imperfections are  shown schematically in Fig. \ref{fig:fab} (a) for periodic poling. 

Impact of imperfections on conversion efficiency was studied previously for periodically poled crystals (e.g random variations in domain walls \cite{Helmfrid:91}, missed domains \cite{Karlsson1996}, and deviations in duty-cycle \cite{Karlsson1997}). Here, we study this for custom poled crystals, and also consider how imperfections affect heralded-photon  spectral purity. To gain information about photon pair generation,  we compute the peak amplitude of the $|JSA|$, and compare it with that generated by a periodically poled crystal.

We consider poled KTP pumped by a pulsed laser, assuming a Gaussian PEF with $\bar\omega_p=\SI{2.38e15}{\hertz}$ and $\sigma_{\textsc{pef}}=\SI{4.5e11}{\hertz}$. Our goal was to generate degenerate photons at $\bar\omega_i=\bar\omega_s=\SI{1.16e15}{\hertz}$, in the symmetric GVM configuration, which corresponded to a coherence length $l_c=23.05~\mu$m (yielding a poling period  $\Lambda=46.1~\mu $m for periodically poled and custom-duty cycle methods, and a domain width equal to $l_c$ for the custom domain orientation method). When generating the JSA, we used a discretization of $N=100$, and a spectral range of $\SI{7.0e12}{\hertz}$, centered around the degenerate frequencies. 

For comparison, we fixed the generated photon bandwidth across all methods. To achieve this, different methods required different crystal lengths. For our simulations, we used: $L=30.5$ mm (1320 domains) for periodic poling; $L=36.9$ mm (1600 domains) for custom duty cycle; and $L=46.1$ mm (2000 domains) for custom domain orientation.

\emph{Over-/under-poling---}Simulations for over- and under-poling are analogous,  we thus restricted our simulations to over-poling.  We systematically increased the widths of flipped segments while proportionally decreasing the unflipped segments. We considered up to a 15\% change in segment width and found that 
the effect was negligible on both purity and peak amplitude for all methods. Deviations of around 5\% were reported in \cite{Karlsson1997}.

\emph{Random variations in wall positions---}We ran Monte Carlo simulations averaged over 100 data points, based on a Gaussian distribution centered around the expected position of each domain, parameterized by the standard deviation $\sigma_{\textsc{rvd}}$. The effect of this error on purity was negligible for all poling methods, even for high errors. The effect on peak amplitude was more pronounced, but consistent across all methods (although, the customized duty-cycle method showed slightly more spreading across simulations). Fig. \ref{fig:fab} (b)  shows those results.

In our simulations, we considered $\sigma_{\textsc{rvd}}$ to range between $0$ and $0.12\times l_c$. Errors in standard crystals made by established manufacturers will typically be on the low end of this range, but others have reported errors of $\sigma_{\textsc{rvd}}=0.08\times l_c$ in  two different experiments involving lithium niobate waveguides  \cite{Pelc:10,Pelc:11}. Furthermore, as future experiments push the boundaries of  what is possible to manufacture, e.g. to implement methods such as those based on sub-coherence length domains  \cite{graffitti2017pure}, tolerance to relatively high errors will be relevant. 

\emph{Missed domains---}We modelled missed domains by considering segments pointing in one direction, then flipping the sign of a randomly selected subset of these.  We average over 100 data points for each \% value. The effect of missed domains on purity was also negligible for all poling methods, even for high errors (although, the customized duty-cycle method showed slightly more spreading across simulations). The effect on peak amplitude was again more pronounced, but consistent across all methods. Fig. \ref{fig:fab} (c)  shows those results.

In summary, the fabrication imperfections considered here impact all poling methods equivalently.  Errors due to over- and under-poling have negligible effects. Errors due to random variations in wall positions and missed domains do impact peak amplitude, but have negligible effect on  heralded photon spectral purity.

\subsubsection{Undesirable PDC generation far from the central PMF peak due to nonlinearity shaping}

The nonlinearity shaping techniques discussed in this section shape the PMF through interference between the PMFs of individual domains. The PMF can be shaped as desired only within a certain spectral range of interest, and outside this range, the nature of interference can generate undesired amplitude.

\begin{figure}[t]\center
\vspace{-1.3cm}
\includegraphics[height=0.62\textwidth]{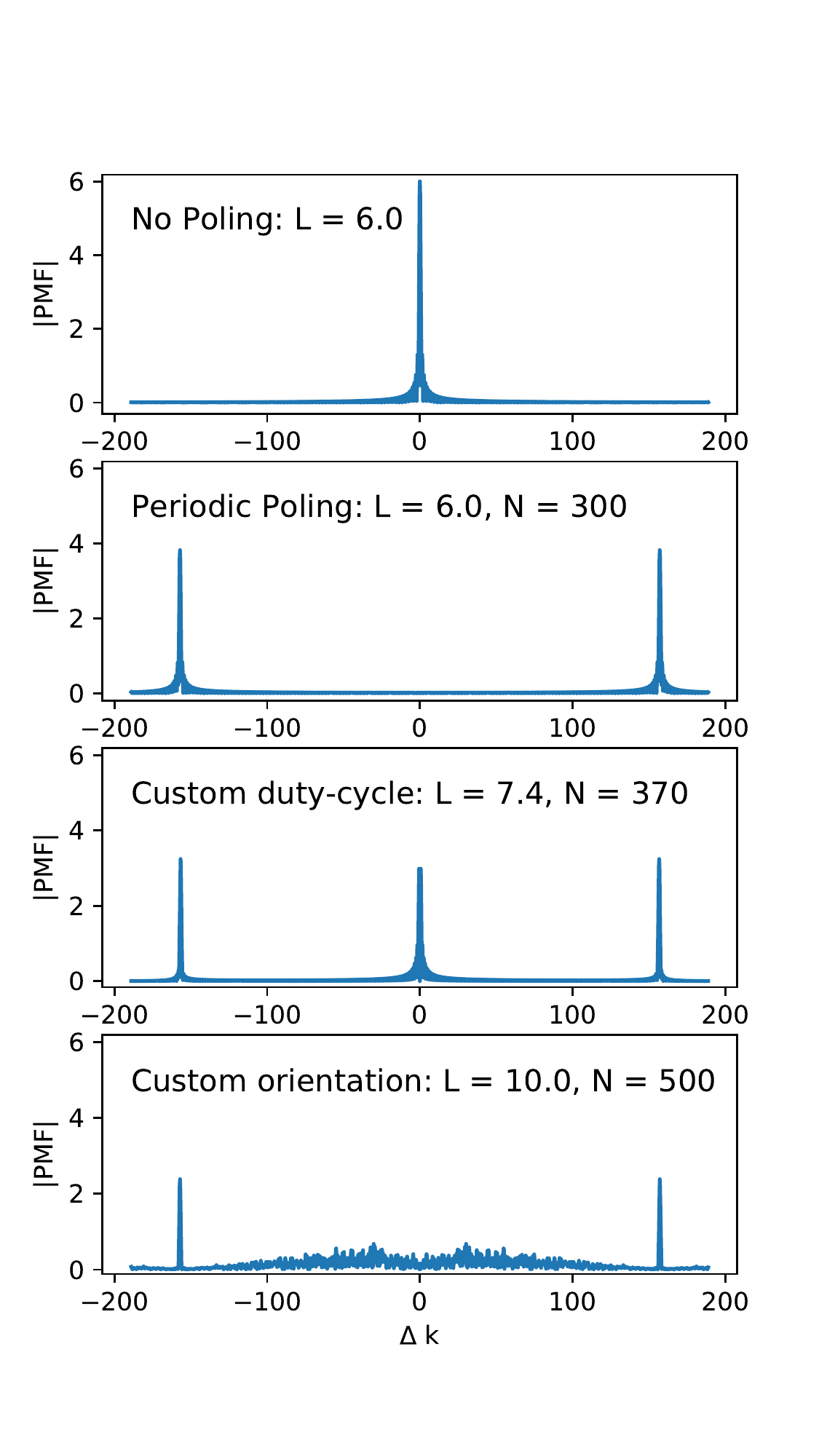}\hspace{-0.4cm}\includegraphics[height=0.62\textwidth]{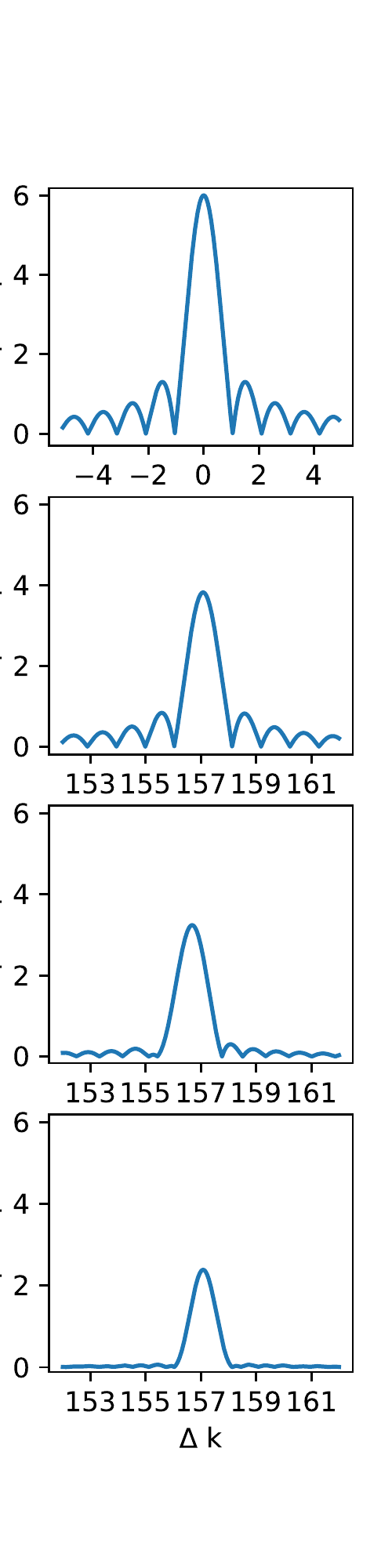}
\vspace{-1.2cm}
\caption{The PMF for an unpoled crystal compared with the PMFs for three poling methods. Undesirable PDC generation far from the
central PMF peak arises outside of the region of interest when poling methods are used. Left: broad $\Delta k$ range showing undesired and desired PDC generation. Right: $\Delta k$ range of interest showing only desired PDC generation. The coherence length $l_c$ was the same for all methods (yielding a poling period  $\Lambda=2l_c $m for periodically poled and custom-duty cycle methods, and a domain width equal to $l_c$ for the custom domain orientation method). For comparison, we fixed the generated photon bandwidth across all methods. To achieve this, different methods required different crystal lengths $L$. $N$ is the number of domains.}
\label{fig:fourier}
\end{figure}

Undesirable PDC generation far from the central PMF peak arises in all poling techniques, but the nature of that amplitude differs.  Fig. \ref{fig:fourier} shows the undesirable PDC generation for three poling patterns, compared with an unpoled crystal. For periodic poling, these regions are concentrated at $\Delta k=\pm n 2\pi/\Lambda$. For the customized duty-cycle method introduced by Dixon et al. \cite{dixon2013spectral}, there is additional amplitude peaked at $\Delta k = 0$. For the customized domain orientation method introduced by Tambasco et al \cite{tambasco2016domain}, the additional amplitude is more spread out. 

For periodic poling and the customized duty-cycle method, the undesirable PDC generation is typically far outside the spectral range of the detectors and therefore gets filtered out automatically. For the customized domain orientation method, the undesired PDC generation is closer to the spectral range of interest, and might need to be filtered out deliberately. This raises the question: if the motivation for nonlinearity shaping was to avoid destruction of photon-number correlations caused by filtering, is nonlinearity shaping a good idea when filtering is required anyway?

The answer lies in the nature of the filtering. Filtering  preserves photon number correlations if the filter is partially transmissive \emph{only} at frequencies for which the JSA has negligible support (the special case of this is a filter described by a top-hat function with unit transmittance, which can be used in any region of the JSA).  If the region containing the undesirable PDC generation is far enough away, from the region of interest, to ensure no overlap between where the JSA has support and where the filter is partially transmissive, then the undesired PDC generation can be safely filtered out without destroying photon-number correlations \cite{branczyk2010optimized,laiho2011testing,christ2012limits,christ2014theory,meyer2017limits}. So as long as the undesired PDC generation is sufficiently far away from the desired PDC generation, nonlinearity shaping is a good idea.

If the undesired PDC generation is too close to the desired PDC generation, it might be possible to suppress its generation by engineering a Bragg grating into the nonlinear material to induce a photonic stop band \cite{Helt2017}.

\section{Inferring the purity of the heralded single photon}\label{sec:JSAJSI}

In the previous section, we considered the  design of a \emph{spectrally pure} heralded single photon source. In this section, we focus on characterizing the  spectral purity of the source once it is built. While the spectral purity of a heralded photon cannot be measured directly, it can be inferred from other measurements, such as those of the JSA, the JSI (in special cases), or the Hong-Ou-Mandel (HOM) visibility \cite{branczyk2017hong}. 

Any experimental measurement of the JSA or JSI necessarily yields a discretized approximation over a finite spectral range. In this section, we study how different discretizations and spectral ranges  impact the inferred spectral purity.

We also study the effect of using the square root of the JSI as a proxy for the JSA (recall that $\mathrm{JSI}=|\mathrm{JSA}|^2$). This is important because many methods that measure the photons' joint spectral properties---such as scanning-monochromators measurements, fibre-spectroscopy techniques or stimulated emission tomography \cite{kim2005measurement,kuzucu2008joint,avenhaus2009fiber,liscidini2013stimulated,eckstein2014high,zielnicki2018joint}---lack spectral phase and sign information. These methods really measure the JSI. We show that using the square root of the JSI for  purity estimation involves some pitfalls if the JSA has phase-correlations, including sign-changes. 

The only method to reconstruct the JSA directly (including phase correlations) is the phase-sensitive stimulated emission tomography \cite{jizan2016phase}, but this is experimentally hard and is not a widespread technique. If there is reason to believe that the JSA has both positive and negative regions, or if it has additional temporal correlations such as those that come from chirped pulses---and it is not possible to do phase-sensitive stimulated emission tomography---then one may do a two-photon HOM interference experiment. We show that the visibility of this experiment predicts the spectral purity even in the case of chirped pulses.

\subsection{Discretization and spectral range}

In the  Section \ref{sec:pur}, we saw that the purity can be calculated from the Schmidt decomposition of the JSA. To do this in practice, the JSA is discretized into frequency bins, over finite ranges of signal and idler frequencies, then represented as a complex-valued matrix.  The Schmidt decomposition is then computed numerically using a singular value decomposition (SVD) \cite{miszczak2011singular,laudenbach2016modelling}  of the matrix representation of the JSA. 

Whether the discretized JSA is obtained experimentally using, e.g. phase-sensitive stimulated emission tomography, or constructed from the analytical form of the JSA, it is crucial to correctly choose the spectral range of  both the signal and the idler photons and the number of frequency-bins used for the discretization. 
In this section, we analyze the effects of discretization and spectral range using a JSA constructed from a sinc-shaped PMF and sech-shaped pump function in the symmetric GVM regime ($\theta=\pi/4$).
 
 \begin{figure}[b]\center
\includegraphics[width=0.45\textwidth]{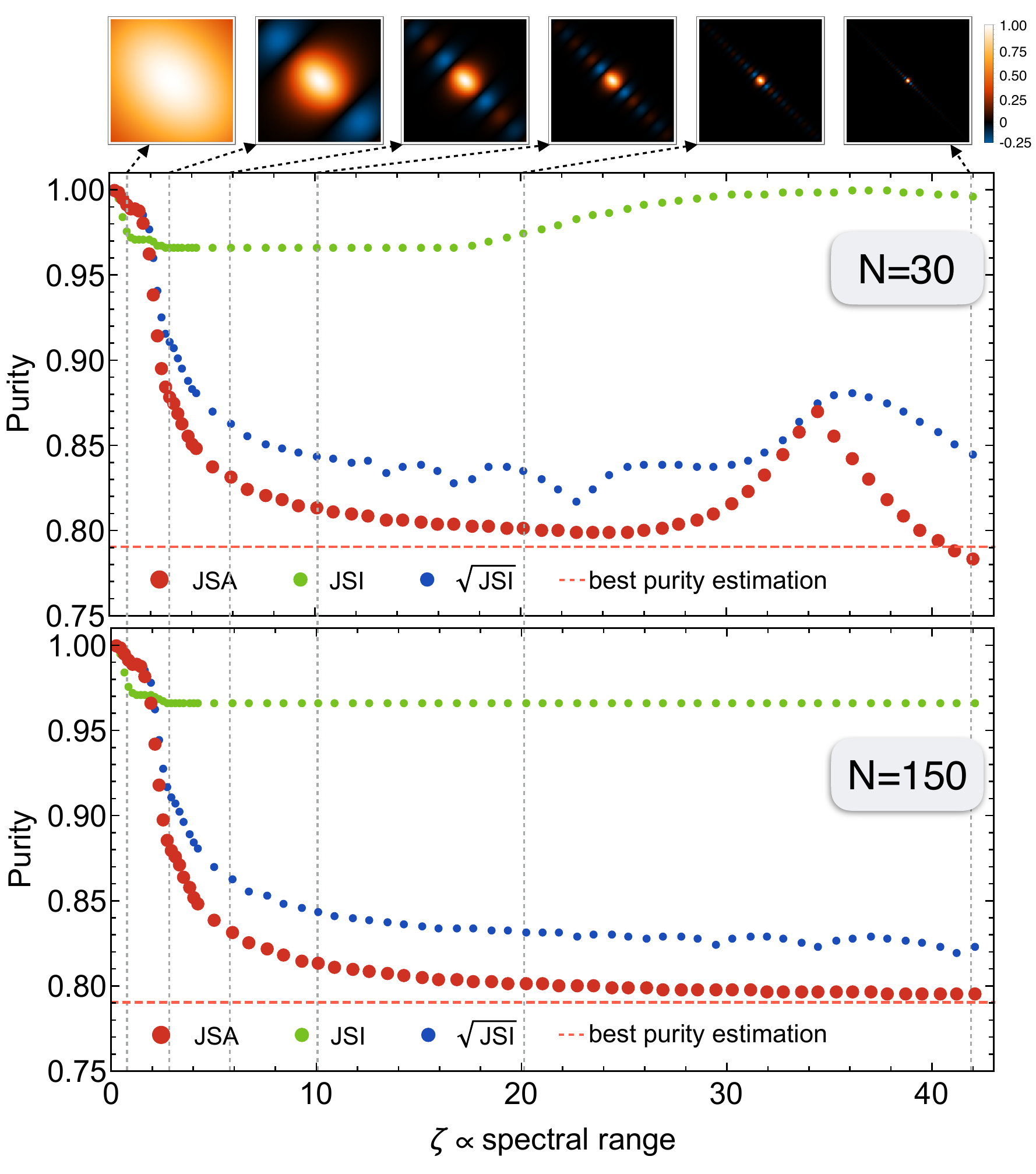}
\caption{
Dependence of the heralded photon spectral purity on the spectral range for two different values of resolution N. Red points correspond to the inferred purity, while green and blue points correspond to an inferred purity-like parameter computed from Eq. \ref{eq:purity} using the JSI or $\sqrt{\textrm{JSI}}$, respectively,  in place of the JSA.
}
\label{fig:spectral_range}
\end{figure}
 
To study the effect of a finite spectral range, we fix the resolution (defined as the number of frequency bins) and construct matrix representations of the JSA for increasing spectral ranges. We parameterize the spectral range by $\zeta$, which is  the ratio between the spectral range used in the JSA calculation and the average PDC photon bandwidth (defined as the FWHM of the marginal spectral distributions of the photons).  We then find the singular values of each matrix and use it to compute the purity according to Eq. (\ref{eq:purity}). We find that for a fixed resolution, there is an optimal value of spectral range. This can be seen in  Fig. \ref{fig:spectral_range} where the red markers show the purity computed from the JSA as a function of spectral range for resolutions of ($N=30$ and $N=150$). Initially, as the spectral range increases, more of the true spectrum is included in the finite representation of the JSA, and the value of the purity approaches the true value. But since the number of frequency bins is fixed, each bin gets larger as the spectral range continues to increase, and  eventually cannot capture detailed features of the JSA, so the inferred purity diverges from the true purity. 

Mathematically, we know that only the SVD of the JSA can yield the actual purity. But since others  (e.g. \cite{mosley2008heralded,edamatsu2011photon,jin2013widely,weston2016efficient,chen2017efficient,zielnicki2018joint}) have used the JSI or the $\sqrt{\text{JSI}}$ ($|\mathrm{JSA}|$) to get information about the purity from experiments, we also construct matrix representations of the JSI and $\sqrt{\mathrm{JSI}}$ ($|\mathrm{JSA}|$) and compute a purity-like parameter using the singular values of these matrices. Fig. \ref{fig:spectral_range} shows that, using this approach, neither the JSI (green) or the  $\sqrt{\mathrm{JSI}}$ (blue) provide  good estimates of the true purity.

\begin{figure}[t]\center
\includegraphics[width=0.45\textwidth]{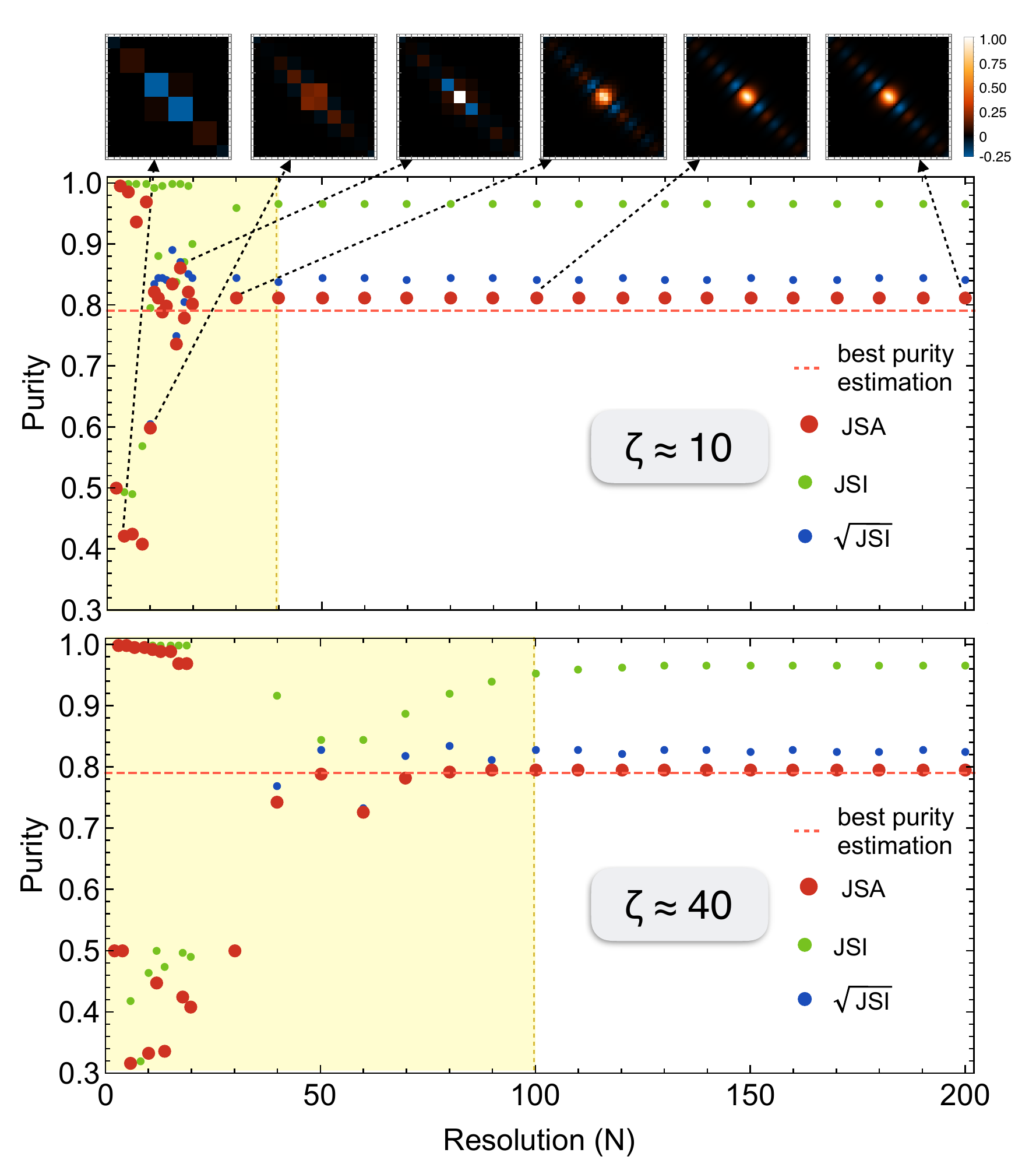}
\caption{
Dependence of the photon purity on the spectral resolution for two different values of spectral range. Red points correspond to the inferred purity, while green and blue points correspond to an inferred purity-like parameter computed from Eq. \ref{eq:purity} using the JSI or $\sqrt{\textrm{JSI}}$, respectively,  in place of the JSA.
}
\label{fig:discretisation}
\end{figure}

To study the effect of discretization, we fix the spectral range and construct matrix representations of the JSA for a range of discretizations. As before, we then find the singular values of each matrix and use it to compute the purity according to Eq. (\ref{eq:purity}). We find that for a fixed spectral range, the purity converges as the discretization is increased. This can be seen in  Fig. \ref{fig:discretisation} where the red markers show the purity computed from the JSA as a function of discretization for spectral range of $\zeta \approx 10$ and $\zeta \approx 40$. Indeed, the calculated spectral purities in the highlighted yellow regions ($N<40$ for the smaller spectral range, $N<100$ for the larger) is extremely sensitive to $N$, and the corresponding resolutions aren't suitable for estimating accurately the spectral properties of the PDC photons. At higher resolutions ($N>40$ and $N>100$) the inferred purities converge to a single value of spectral purity. 

As before, we also construct matrix representations of the JSI and $\sqrt{\mathrm{JSI}}$ ($|\mathrm{JSA}|$) and compute  purity-like parameters using the singular values, shown in Fig. \ref{fig:discretisation}. Both converge, but to the wrong value, thus neither provide very good estimates of the true purity.
This discrepancy is due to the sinc shaped PMF having both positive and negative amplitude components.  For ideal Gaussian-shaped PEF and PMF, the purity-like parameter  would converge to the true purity.

\begin{figure}[tb]\center
\includegraphics[width=0.45\textwidth]{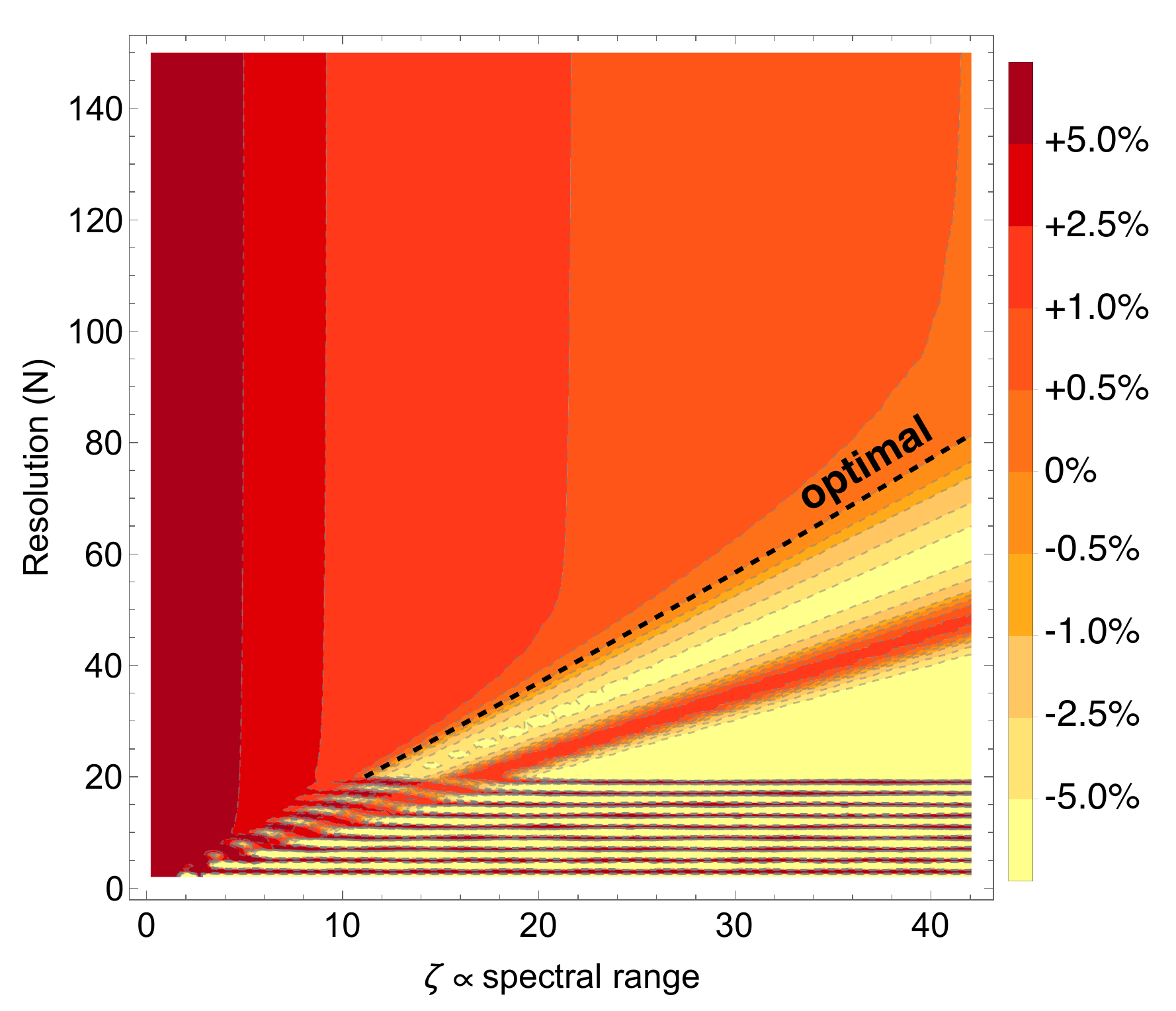}
\caption{
Contour plot of the single-photon purity vs spectral range $\zeta$ and discretisation resolution $N$ for a sech pulse group-velocity matched with a sinc-shaped PMF.
Different colours correspond to different ranges of purity respect to its most accurate estimate (0.79 -- computed with $\zeta\approx630$ and $N=3000$). The black dashed line represents the exact purity value.
}
\label{fig:range_discretisation}
\end{figure}

To study the interplay between the discretization and the spectral range, 
we compute the JSA separability at different discretizations and spectral ranges, and we compare it with a very accurate purity estimate obtained via SVD from a JSA with $\zeta\approx630$ and $N=3000$. 
We show the results in Fig. \ref{fig:range_discretisation}. The purity is significantly overestimated for small spectral ranges $\zeta < 10$, while a coarse discretization ($N<20$) leads to noisy results. In general, reliable purity values are obtained in the top-right corner of the plot.

\begin{figure}[htb]\center
\includegraphics[width=0.45\textwidth]{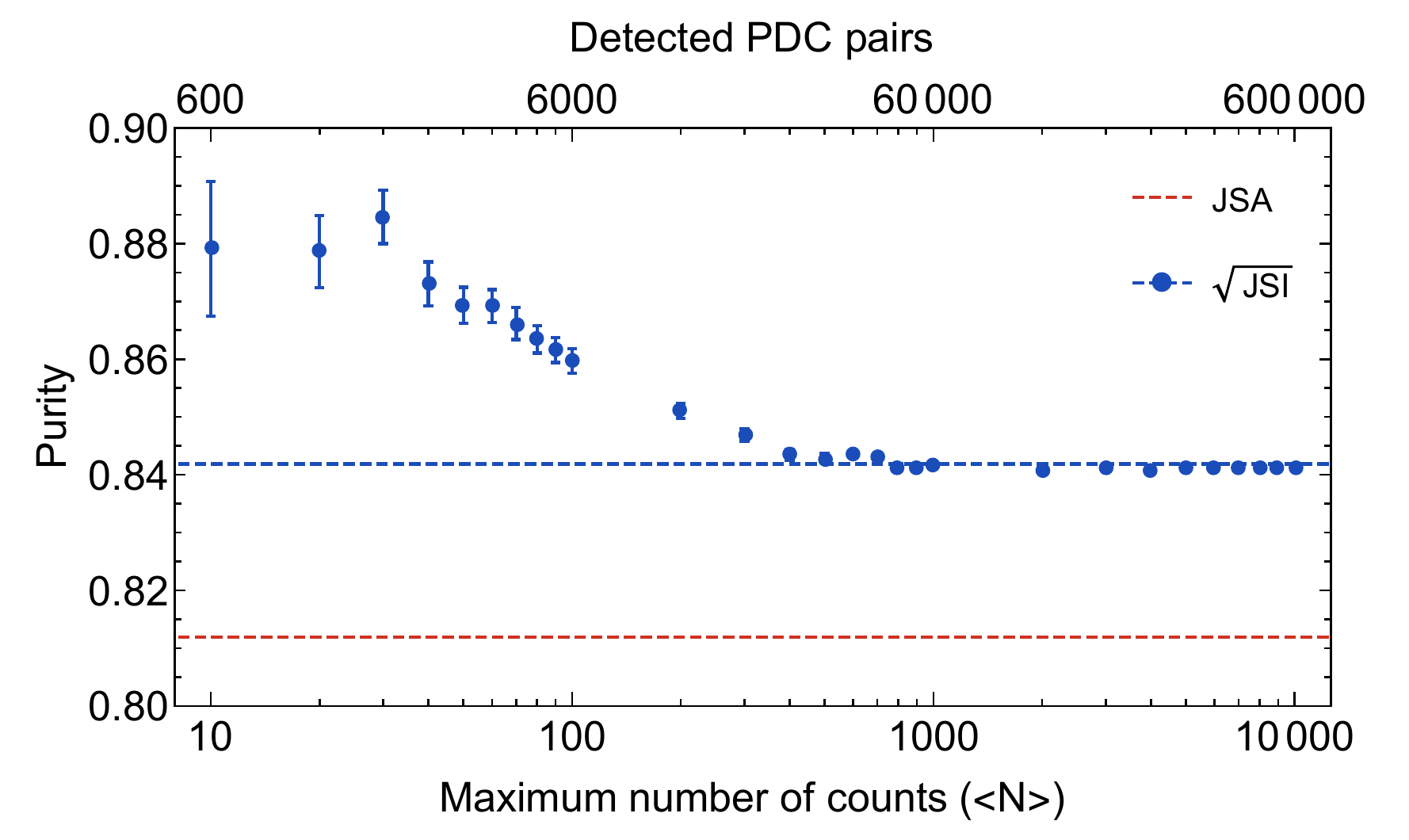}
\caption{
Monte Carlo simulation of the purity-like parameter computed from Eq. \ref{eq:purity} using the $\sqrt{\textrm{JSI}}$  in place of the JSA. On the x-axis the maximum number of counts in the ``brightest'' frequency bin or, equivalently, the total number of detected PDC pairs. Each data point is the mean of 1000 simulated samples (enough for the convergence of the algorithm), while the error bars are the standard deviations. The dashed lines are the  purity simulated from the actual JSA (red) and the  purity-like parameter simulated from the square root of JSI (blue).
}
\label{fig:JSImontecarlo}
\end{figure}

We finally show what happens in an actual experiment when the JSI is measured with limited statistics, i.e. detecting a finite number of coincident photons for each frequency bin according to the spectral probability distribution of the bi-photon state. This is an important detail since bi-photon spectroscopy (via, e.g.,  scanning-monochromator  or  fibre-spectroscopy techniques \cite{zielnicki2018joint}) can be very lossy and return very low count rates anywhere but in the central frequency bin.
Again, we analyse the case of a sinc-shaped PMF matched with a sech pulse in symmetric GVM condition, we we consider a $100 \times 100$ JSI and $\zeta \approx 10$.
We perform a Monte Carlo simulation assuming Poissonian distribution of the detected coincidences.
The results are shown in Fig. \ref{fig:JSImontecarlo}. We find that the estimated purity-like parameter converges to the expected value when the average number of PDC pairs detected in the ``brightest'' frequency bin is approximatively $1000$  (equivalently, for $60000$ overall detected PDC pairs).

We conclude that estimating purity from joint spectral measurements has a number of pitfalls. Measurements based on the JSA and the JSI are impacted by limited spectral range and rough discretization because of limited spectral resolution. Measurements based on the JSI are further impacted by finite photon-counting statistics. In the case of the JSI, even if the characterization is carried out meticulously, the purity-like parameter inferred from the SVD (which, as discussed above, can sometimes correspond to the spectral purity) is at best a rather loose upper bound. Most experiments in the literature which computed the purity-like parameter from JSI measurements (e.g. \cite{mosley2008heralded,gerrits2011generation,yabuno2012four,jin2013widely,harder2013optimized,weston2016efficient,francis2016all,allgaier2017fast,allgaier2017highly,chen2017efficient,zielnicki2018joint,greganti2018tuning}) therefore may have over-estimated its value. 
There are, however, also examples of good (but not yet optimal) practice \cite{fang2014fast,gerrits2015spectral,meyer2017limits}.

\subsection{Inferring the purity via  two-photon interference} \label{sec:HOM}

A more reliable benchmark for  heralded-photon spectral purity is the beamsplitter (BS) interference visibility between two identical photons  \cite{graffitti2018independent}. If the two interfering photons are pure and indistinguishable they exit the BS from the same output mode. If they are either not pure or distinguishable (or both),  they don't interfere perfectly and can exit the BS from both ports simultaneously. This is quantified by the visibility:
\begin{align}
V=1-\frac{N_{\text{min}}}{N_{\text{max}}}\,,
\end{align}
where  $N_{\text{max}}$ is the number of photon pairs that exit the BS from opposite ports after arriving at the BS simultaneously, while  $N_{\text{min}}$ is the number of photon pairs that exit the BS from opposite ports after arriving  at the BS at different times for identical photons. The visibility is equal to the heralded-photon spectral purity \cite{branczyk2017hong}.

\begin{figure}[htb]\center
\includegraphics[width=0.48\textwidth]{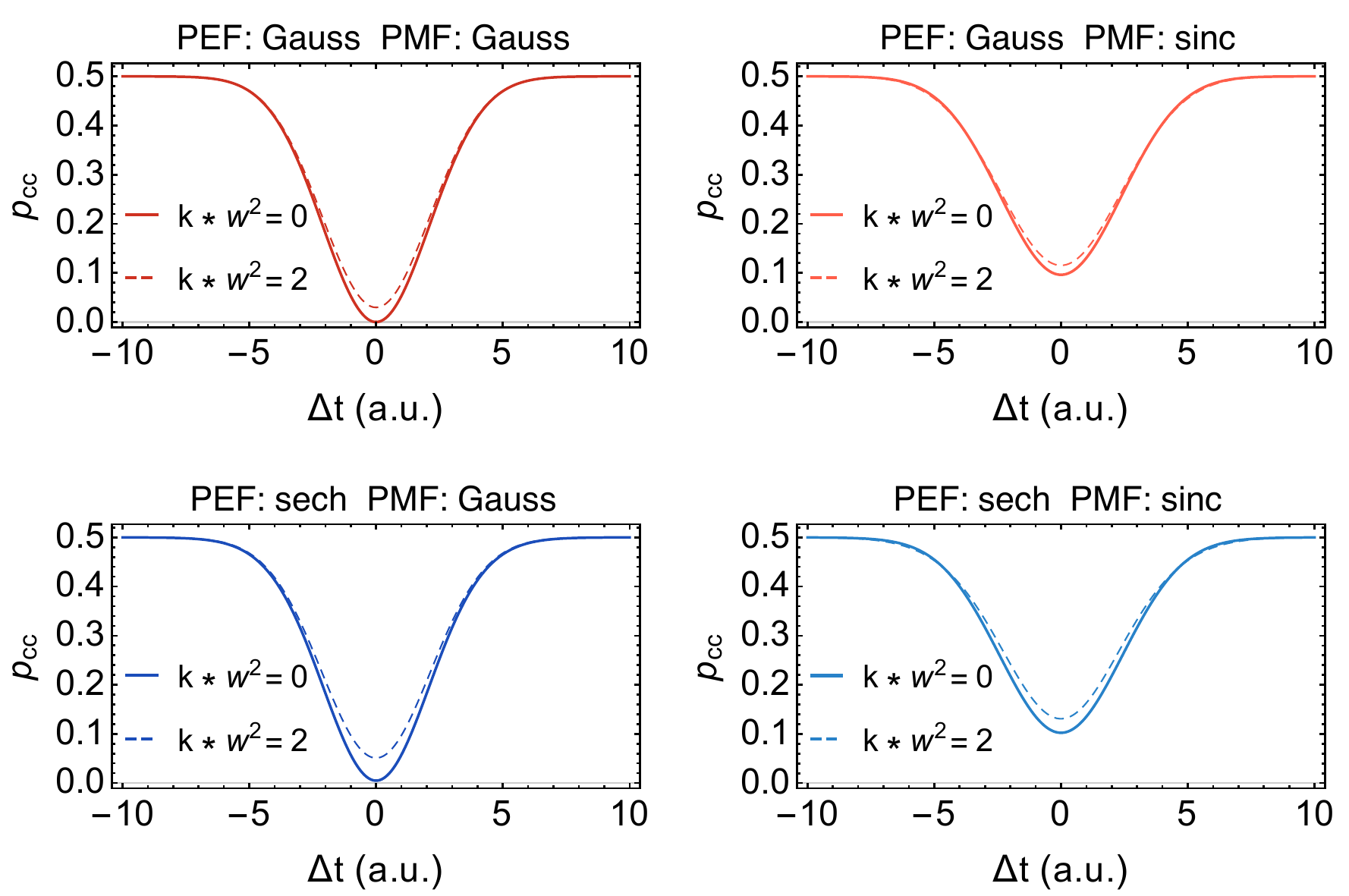}
\caption{
Two-photon interference patterns for four combinations of PEF and PMF shapes in symmetric GVM condition. Both a transform-limited case and a $k w^2 =2$ case are considered.
}
\label{fig:HOM}
\end{figure}

Fig. \ref{fig:HOM} shows that the two-photon interference is sensitive to the phase information of the JSA: the visibilities of the interference patterns match the purities obtained via Schmidt decomposition shown in Fig. \ref{fig:chirped_purity}.

If there is reason to believe that the JSA has both positive and negative regions, or if it has additional temporal correlations such as those that come from chirped pulses---and it is not possible to measure the JSA---then  a two-photon HOM interference experiment is a good option to infer the spectral purity.

\section{Conclusion}
We investigated a number of practical issues relevant to the design and characterisation of single-photon sources based on parametric downconversion in a group-velocity-matched regime. 

We showed that when realistic laser pulses and realistic nonlinear crystals are used, the pulse laser and PDC bandwidths (i.e. choice of crystal length as a function of pulse shape and duration)  that optimize heralded photon spectral purity, differ to those previously found for ideal Gaussian functions. We highlighted the existence of unwanted PDC generation that arises from different nonlinearity shaping methods. We also considered fabrication imperfections and found that while they did impact conversion efficiency, the impact on heralded photon spectral purity was negligible.

We examined state characterization methods based on the joint spectrum of bi-photons or two-photon interference. We found that discretization and spectral range of the joint spectrum played a large role in correctly inferring the heralded photon spectral purity. We also showed that in cases where the joint spectral amplitude changes sign or contains non-trivial phases, inferring the purity from  the joint spectral intensity leads to incorrect results. We showed that in those cases, if it is not possible to measure the joint spectral amplitude, then  a two-photon HOM interference experiment is a good option to infer the spectral purity.

The theory developed in this paper is for PDC in $\chi^{(2)}$ materials, but our analysis on how the PEF shape and chirp impact the bi-photon properties can be extended to Four Wave Mixing in $\chi^{(3)}$ materials, which are a building block of integrated LOQC. Furthermore, our results on JSA characterization apply directly to bi-photons generated via FWM sources. 

The sum of these considerations provide a recipe for the correct choice of: the experimental parameters for matching pulse laser to PDC bandwidths; the optimal approach to nonlinearity tailoring; and the parameters for characterising the purity of the resulting photons. Taking these considerations into account will further improve the quality of PDC photon sources in terms of brightness, spectral purity, and heralding efficiency. We therefore expect our results to be of practical interest to researchers building the next generation of nonlinear sources of separable photon pairs.

\section*{Funding Information}
This work was supported by the UK Engineering and Physical Sciences Research Council (grant number EP/N002962/1). F.G. acknowledges studentship funding from EPSRC under grant no. EP/L015110/1. Research at Perimeter Institute is supported by the Government of Canada through Industry Canada and by the Province of Ontario through the Ministry of Research and Innovation. We acknowledge the support of the Natural Sciences and Engineering Research Council of Canada  (funding reference number RGPIN-2016-04135).


\bibliography{bibliography_Practicalities_PDC}{}

\end{document}